\documentclass[%
 aps,prl,twocolumn,superscriptaddress,
 amsmath,amssymb,amsfonts,
 floatfix
]{revtex4-2}
\usepackage[T1]{fontenc}
\usepackage{lmodern}
\usepackage{microtype}
\usepackage{graphicx}
\usepackage{bm}
\usepackage[dvipsnames]{xcolor}
\usepackage{hyperref}
\hypersetup{
    colorlinks=true,
    citecolor=Green,
    linkcolor=Red,
    urlcolor=NavyBlue
}
\usepackage[capitalise]{cleveref}
\usepackage{soul}
\usepackage{mathtools}
\usepackage{siunitx}

\let\oldhat\hat
\renewcommand{\vec}[1]{\mathbf{#1}}
\renewcommand{\hat}[1]{\mathbf{\oldhat{#1}}}

\DeclarePairedDelimiter\abs{\lvert}{\rvert}

\begin{abstract}
 In light of recent experimental data indicating a substantial thermal Hall effect in square lattice antiferromagnetic Mott insulators, we investigate whether a simple Mott insulator can sustain a finite thermal Hall effect. We verify that the answer is ``no'' if one performs calculations within a spin-only low-energy effective spin model with non-interacting magnons. However, by performing determinant quantum Monte Carlo simulations, we show the single-band $t$-$t'$-$U$ Hubbard model coupled to an orbital magnetic field does support a finite thermal Hall effect when $t' \neq 0$ and $B \neq 0$ in the Mott insulating phase. We argue that the (carrier agnostic) necessary conditions for observing a finite thermal Hall effect are time-reversal and particle-hole symmetry breaking. By considering magnon-magnon scattering using a semi-classical Boltzmann analysis, we illustrate a physical mechanism by which finite transverse thermal conductivity may arise, consistent with our symmetry argument and numerical results. Our results contradict the conventional wisdom that square and triangular lattices with SU(2) symmetry do not support a finite thermal Hall effect and call for a critical re-examination of thermal Hall effect data in insulating magnets, as the magnon contribution should not be excluded \emph{a priori}.
\end{abstract}

\begin{document} 

\title{Intrinsic Thermal Hall Effect in Mott Insulators}

\author{Jixun K. Ding}
\thanks{These authors contributed equally to this work}
\affiliation{Stanford Institute for Materials and Energy Sciences,
SLAC National Accelerator Laboratory, 2575 Sand Hill Road, Menlo Park, CA 94025, USA}
\affiliation{Department of Applied Physics, Stanford University, Stanford, CA 94305, USA}

\author{Emily Z. Zhang}
\thanks{These authors contributed equally to this work}
\affiliation{Department of Physics, University of Toronto, Toronto, Ontario M5S 1A7, Canada}

\author{Wen O. Wang}
\affiliation{Stanford Institute for Materials and Energy Sciences,
SLAC National Accelerator Laboratory, 2575 Sand Hill Road, Menlo Park, CA 94025, USA}
\affiliation{Department of Applied Physics, Stanford University, Stanford, CA 94305, USA}

\author{Tessa Cookmeyer}
\affiliation{Kavli Institute for Theoretical Physics, University of California, Santa Barbara, CA 93106, USA}

\author{Brian Moritz}
\affiliation{Stanford Institute for Materials and Energy Sciences,
SLAC National Accelerator Laboratory, 2575 Sand Hill Road, Menlo Park, CA 94025, USA}

\author{Yong Baek Kim}
\email{yongbaek.kim@utoronto.ca}
\affiliation{Department of Physics, University of Toronto, Toronto, Ontario M5S 1A7, Canada}

\author{Thomas P. Devereaux}
\email{tpd@stanford.edu}
\affiliation{Stanford Institute for Materials and Energy Sciences,
SLAC National Accelerator Laboratory, 2575 Sand Hill Road, Menlo Park, CA 94025, USA}
\affiliation{
Department of Materials Science and Engineering, Stanford University, Stanford, CA 94305, USA}
\date{\today}

\maketitle 

\textit{Introduction.}--- Thermal Hall transport is a powerful experimental probe for diagnosing the underlying excitations in quantum materials. At its core, transverse heat transport is sensitive to the nontrivial topology of the heat carriers, and the mechanism by which the carriers acquire this topology depends on the nature of the excitation. This sensitivity is particularly advantageous in insulators where conventional charge transport experiments cannot be performed, allowing for the detection of charge neutral excitations, such as magnetic fluctuations (magnons)~\cite{McClarty2022}, lattice vibrations (phonons)~\cite{Qin2012, Saito2019}, and spin fractionalization (spinons)~\cite{Katsura2010}. Understanding the unique signatures stemming from various quasiparticle excitations and the interplay between them continues to be a persistent pursuit within the field. 

Recently, a large, negative thermal Hall conductivity $\kappa_{xy}$ was measured in the undoped Mott insulating phase of various cuprate superconductors~\cite{Grissonnanche2019,Grissonnanche2020,Boulanger2020}, as well as the antiferromagnetic insulator Cu$_3$TeO$_6$~\cite{Chen2022e}. While the large signal observed down to low temperatures has mainly been attributed to phonons, the exact mechanism by which the phonons acquire chirality remains under debate~\cite{Sun2022,Flebus2022,Ye2021,Guo2022,Mangeolle2022,Oh2024}. This uncertainty leads us to consider other charge-neutral excitations that yield a finite thermal Hall signal. 

A sizable $\kappa_{xy}$ can also arise from topological magnon excitations of magnetically ordered states~\cite{Onose2010,Matsumoto2014,Murakami2017,McClarty2022,Zhang2024}. The magnons may generically acquire a nontrivial topology due to exchange interactions that break global spin rotation (SU(2)) symmetry, such as the bond-dependent Kitaev interaction on honeycomb lattices ~\cite{McClarty2018,Zhang2021}, or the antisymmetric Dzyaloshinskii-Moriya (DM) interaction~\cite{Doki2018,Laurell2018}, leading to finite transverse thermal transport. In the case where SU(2) symmetry is preserved, e.g. with Heisenberg-type Hamiltonians with a ring exchange, certain lattice geometries are believed to be barred from exhibiting a thermal Hall effect due to various no-go theorems~\cite{Katsura2010}. The antiferromagnetic order on the square lattice seen in cuprate insulators is an example of one of these forbidden geometries. These no-go theorems, however, were derived using non-interacting magnons within the context of linear spin-wave theory (LSWT). While it has been postulated that $\kappa_{xy}$ should still be negligible when perturbing away from this limit (e.g. via a small canting of the spin moments)~\cite{Samajdar2019}, the effects of incorporating magnon-magnon interactions are unknown, but can in principle contribute to the thermal Hall effect~\cite{Carnahan2021,Mook2021,Mook2023,Chatzichrysafis2024}. Moreover, for insulating phases close to the Mott transition, enhanced charge fluctuations leading to higher-order exchange terms may arise in the spin Hamiltonian, whose inclusion may also contribute to thermal Hall transport. Calculating the magnon thermal Hall coefficient including these effects without using simplifying assumptions or perturbative approaches is theoretically challenging, even on the simple square lattice. In an effort to overcome this hurdle, we raise a more fundamental question: Without assuming a particular magnon model, what are the requirements for observing a finite thermal Hall effect in Mott insulators? 

In this paper, we study the $t$-$t'$-$U$ Hubbard-Hofstadter model on the square lattice, which captures essential features of high-$T_c$ cuprates under a magnetic field. We emphasize that this model does not include extrinsic sources of scattering, such as phonons and disorder, so any mechanism for finite transport are intrinsic to the model electonic Hamiltonian. First, we examine the symmetry requirements for a finite thermal Hall response to exist. While our model does not break spin SU(2) symmetry, the applied orbital magnetic field breaks time-reversal symmetry (TRS), and the inclusion of second nearest-neighbor hopping $t'$ breaks particle-hole symmetry (PHS). We demonstrate that one cannot obtain a finite thermal Hall conductivity in particle-hole symmetric systems, even if TRS is broken. Heuristically, we then expect that in systems without PHS, the thermal Hall conductivity is generically nonzero. We use determinant Quantum Monte Carlo (DQMC)~\cite{White1989,Loh1990} to compute $\kappa_{xy}$ in the undoped Mott insulating phase, and explicitly demonstrate the relationship between PHS breaking and nonzero $\kappa_{xy}$. By increasing $\abs{t'}$, which controls the degree of PHS breaking, we observe an increase in the thermal Hall conductivity. We find that the computed $\kappa_{xy}/T$ is on the order of $0.01$ to $0.1\ k_B^2/\hbar$ at $T/t = 1/5, B=0.0625\Phi_0/a^2$, where $\Phi_0=hc/e$ is the magnetic flux quantum and $a$ is the lattice constant.

Finally, we consider magnon-magnon scattering as one possible mechanism for generating a finite thermal Hall effect using the semiclassical theory of Ref. \cite{Chatzichrysafis2024}. Projecting into the low-energy spin Hamiltonian results in an effective $J_1$-$J_2$-$J_\chi$ model, in which the chiral $J_\chi$ term only appears when both $t'$ and $B$ are present. Using a semi-classical Boltzmann analysis, we demonstrate that only the collision processes mediated by the $J_\chi$ interaction contribute to the transverse thermal conductivity, consistent with both our symmetry argument and our numerical results. Our findings illustrate the limitations of linear spin wave theory, and imply that intrinsic spin fluctuations should not be excluded as transverse heat carriers \emph{a priori} when interpreting thermal Hall experiments in insulating magnets.

\textit{Model and Numerical Methods.}---
We study the single-band Hubbard-Hofstadter model
\begin{align}
H &= -\sum_{ ij  \sigma} t_{ij}  \left\{\exp\left[\mathrm{i}\varphi_{ij}\right]c_{i \sigma}^\dagger c_{j \sigma} + \mathrm{h.c.}\right\}  \nonumber \\
&- \mu \sum_{i \sigma} n_{i\sigma} 
+U\sum_{i}\left(n_{i\uparrow} - 1/2 \right)\left(n_{i\downarrow} - 1/2\right), \label{eq:hamiltonian}
\end{align}
on a two-dimensional square lattice at half-filling. The hopping integral $t_{ij} = t$ between nearest neighbor sites $\langle i j\rangle$,  $t_{ij} = t'$ between next-nearest neighbor sites $\langle \langle i j\rangle \rangle $, and $t_{ij} = 0$ otherwise. $\mu$ is the chemical potential, and $U$ is the on-site Coulomb interaction strength. $c_{i\sigma}^{\dagger}$ ($c_{i\sigma}$) is the creation (annihilation) operator for an electron on site $i$ with spin $\sigma=\uparrow,\downarrow$ and $n_{i\sigma} =  c_{i\sigma}^\dagger c_{i\sigma}$ measures the number of electrons of spin $\sigma$ on site $i$. 
A spatially uniform and static orbital magnetic field is introduced by Peierls substitution via the phase
\begin{equation}
\varphi_{ij} = \dfrac{2\pi}{\Phi_0} \int_{\vec{r}_i}^{\vec{r}_j} \vec{A}\cdot d\bm{\ell}, \label{eq:peierls-phase}
\end{equation}
where 
$\vec{r}_i = (r_{ix}, r_{iy})$ is the position of site $i$, and the path integral is taken along the shortest straight line path between sites $i$ and $j$. The vector potential $\vec{A}$ generates the out-of-plane magnetic field $\vec{B} = B \hat{z}$. In this work we use the symmetric gauge $\mathbf{A}(\vec{r})= B(-r_y\hat{x} + r_x\hat{y})/2$.

DQMC simulations of~\cref{eq:hamiltonian} are performed on a finite cluster with lattice constant $a=1$, and $N_x = 8$ and $N_y = 8$ sites in the $\hat{x}$ and $\hat{y}$ directions, respectively. We implement modified periodic boundary conditions~\cite{Assaad2002}, described in detail elsewhere~\cite{Ding2024}.
Requiring that the wave function be single-valued on the torus gives the flux quantization condition 
$\Phi/\Phi_0 = N_\phi/N$, where $N = N_x N_y$ denotes the total number of sites, $\Phi=Ba^2$ is the magnetic flux threading each unit cell, and $N_\phi$ is an integer. Detailed DQMC simulation parameters are listed in Supplementary Materials (SM)~\cite{SuppMats}, Section S1.

Within DQMC simulations, we measure unequal imaginary time electric/heat current - electric/heat current correlation functions, which are related to frequency-resolved transport coefficients via the Kubo formulas~\cite{Shastry2008}. DQMC measures
\begin{equation}
\chi_{\mu\nu,\alpha\beta}(\tau) =\frac{1}{V}\langle J_{\mu,\alpha}(\tau) J_{\nu,\beta}(0)\rangle, \label{eq:current-corr}
\end{equation}
where $\tau$ is imaginary time, $\mu,\nu$ index current type $1,2$ representing charge and heat current respectively, and $\alpha,\beta$ index directions $x,y$. When $\mu=\nu$ and $\alpha = \beta$, MaxEnt analytic continuation~\cite{Jarrell1996} is used to convert correlators in imaginary time, \cref{eq:current-corr}, to retarded correlators in real frequency, $\chi_{\mu\nu,\alpha\beta}(\omega)$, which is proportional to the conductivity. However, when $\mu=\nu$ and $\alpha \neq \beta$, the off-diagonal correlator $\chi_{\mu\mu,xy}(\omega) $ need not be positive over all frequencies, which precludes us from directly applying the standard MaxEnt algorithm~\cite{Reymbaut2015,Reymbaut2017,Fei2021}. This is a known issue for off-diagonal spectral functions, and in this work, we adopt two different methods to circumvent this difficulty and estimate the thermal Hall response: 1) a subtraction method suggested by Ref.~\cite{Reymbaut2015,Reymbaut2017}, involving performing analytic continuation on a composite object, then subtracting off the diagonal component; and 2) a finite-Matsubara-frequency proxy proposed in our earlier work~\cite{Wang2021}, involving estimating the Hall coefficient with its value at the first nonzero Matsubara frequency. We demonstrate that these two methods give qualitatively similar results. A detailed description of these two approaches can be found in SM~\cite{SuppMats}, Section S4. 

\textit{Symmetry Argument.}--- Here, we outline a simple derivation demonstrating that the thermal Hall coefficient is symmetry enforced to be strictly zero when the Hamiltonian respects chiral symmetry (i.e. when $t'=0$ in the square lattice case), even when a nonzero magnetic field breaks TRS. 

Consider the unitary charge conjugation transform $\mathcal{C}$, which acts as $c_{i\sigma} \rightarrow (-1)^i c_{i\sigma}^\dagger, c_{i\sigma}^\dagger \rightarrow (-1)^i c_{i\sigma}$, where $(-1)^i$ depends on the sublattice~\footnote{This is usually called a particle-hole transformation for the Hubbard model at zero applied magnetic field.}. Also consider the anti-unitary time-reversal transformation $\mathcal{T}$, which acts as $i \rightarrow -i$. The combination $\mathcal{C} \cdot \mathcal{T}$ is the anti-unitary chiral symmetry, $\mathcal{S}$~\cite{Chiu2016}. In the presence of a magnetic field, the Hamiltonian \cref{eq:hamiltonian} with $t'/t = 0$ on a square lattice at half-filling breaks $\mathcal{C}$ and $\mathcal{T}$ individually, but satisfies the combined $\mathcal{S}$ symmetry. Considering the $\mathcal{C}$ transformation alone, the Hamiltonian transforms as
\begin{align}
    H(\vec{A}) \quad \rightarrow \quad  \mathcal{C} H(\vec{A}) \mathcal{C}^{-1} = H(-\vec{A}) \label{eq:phs-condition}
\end{align}
while the heat current operator $\mathbf{J}_Q$ transforms as 
\begin{align}
    \vec{J}_Q(\vec{A}) \quad \rightarrow \quad  \mathcal{C} \vec{J}_Q(\vec{A}) \mathcal{C}^{-1} = \vec{J}_Q(-\vec{A}). \label{eq:phs-condition-2}
\end{align}
As a result, the transverse current-current correlator in imaginary time, $\chi_{22,xy}=\frac{1}{V}\langle J_{Q,x}(\vec{A,\tau})J_{Q,y}(\vec{A})\rangle$, and therefore $\kappa_{xy}$, satisfy
\begin{align}
    \chi_{22,xy}(\vec{A}) &= \chi_{22,xy}(-\vec{A}) \\
    \kappa_{xy}(\vec{A})  &= \kappa_{xy}(-\vec{A}).
\end{align}
However, the transport response coefficients must also obey the Onsager-Casimir relation $\kappa_{xy}(\vec{A})=\kappa_{yx}(-\vec{A})$, thus enforcing $\kappa_{xy}(\vec{A})=\kappa_{xy}(-\vec{A})=0$. The same proof also applies to the electrical Hall and Seebeck coefficients, as shown in the SM~\cite{SuppMats}, Section S6. 

On the other hand, when $t'\neq 0$, \cref{eq:phs-condition,eq:phs-condition-2} are no longer satisfied, so the thermal Hall conductivity is symmetry-allowed to be nonzero. Indeed, our numerical findings in the following section not only demonstrate this effect, but that $\abs{t'}$, which controls the degree of PHS breaking, also dictates the magnitude of $\kappa_{xy}$.

\begin{figure}[htbp]
    \centering
    \includegraphics[width=\linewidth]{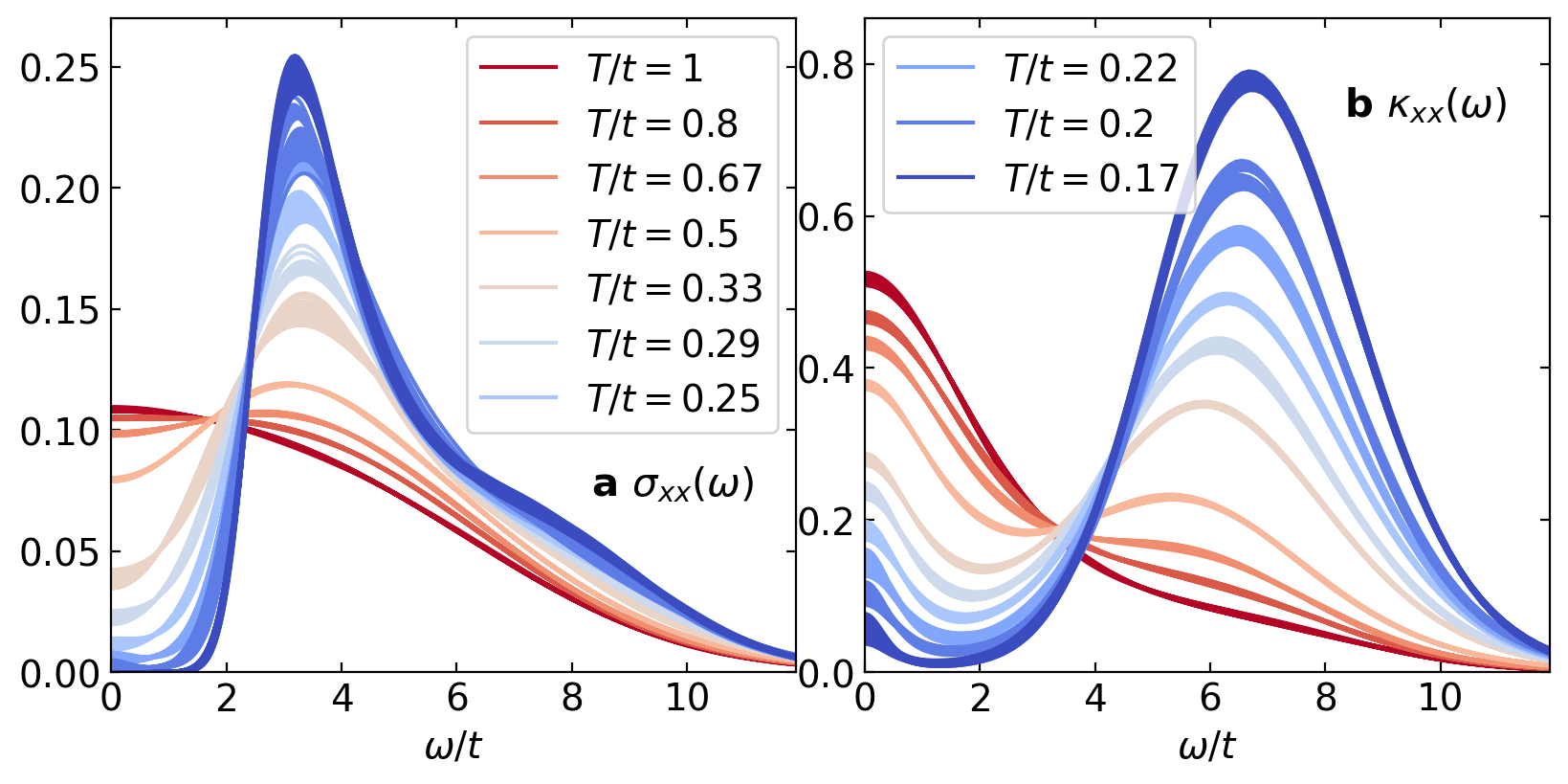}
    \caption[Temperature dependence of longitudinal frequency-dependent electrical and thermal conductivities]{Temperature dependence of longitudinal frequency-dependent \textbf{a} electrical conductivity $\sigma_{xx}(\omega)$, and \textbf{b} thermal conductivity $\kappa_{xx}(\omega)$, for the Hubbard-Hofstadter model with $U/t = 6$, $t'/t=-0.1$, at half-filling $\langle n \rangle = 1$ and fixed field strength $\Phi/\Phi_0 = 4/64$. 100 bootstrap resamples are shown. Both panels share the same legend.}
    \label{fig:optical-T-dep}
\end{figure}

\textit{Results.}--- In \cref{fig:optical-T-dep}, we show representative temperature dependence of longitudinal frequency-dependent electrical and thermal conductivities for Hubbard interaction strength $U/t = 6$, obtained by DQMC simulations. As temperature is lowered, the electrical conductivity becomes gapped, while the thermal conductivity exhibits a drude-like peak near $\omega=0$. This phenomenology is consistent with prior work~\cite{Huang2019a,Wang2022}, and tells us that below temperature scale $T\sim J\sim 4t^2/U$, charge degrees of freedom are frozen out, and magnons are the dominant heat carriers in the system. The behavior shown in \cref{fig:optical-T-dep} is representative in the sense that it does not depend on different next-nearest neighbor hopping $t'/t$ and field strength $B$, as shown in the SM~\cite{SuppMats}, Figure S6.

\begin{figure}[htbp]
    \centering
    \includegraphics[width=\linewidth]{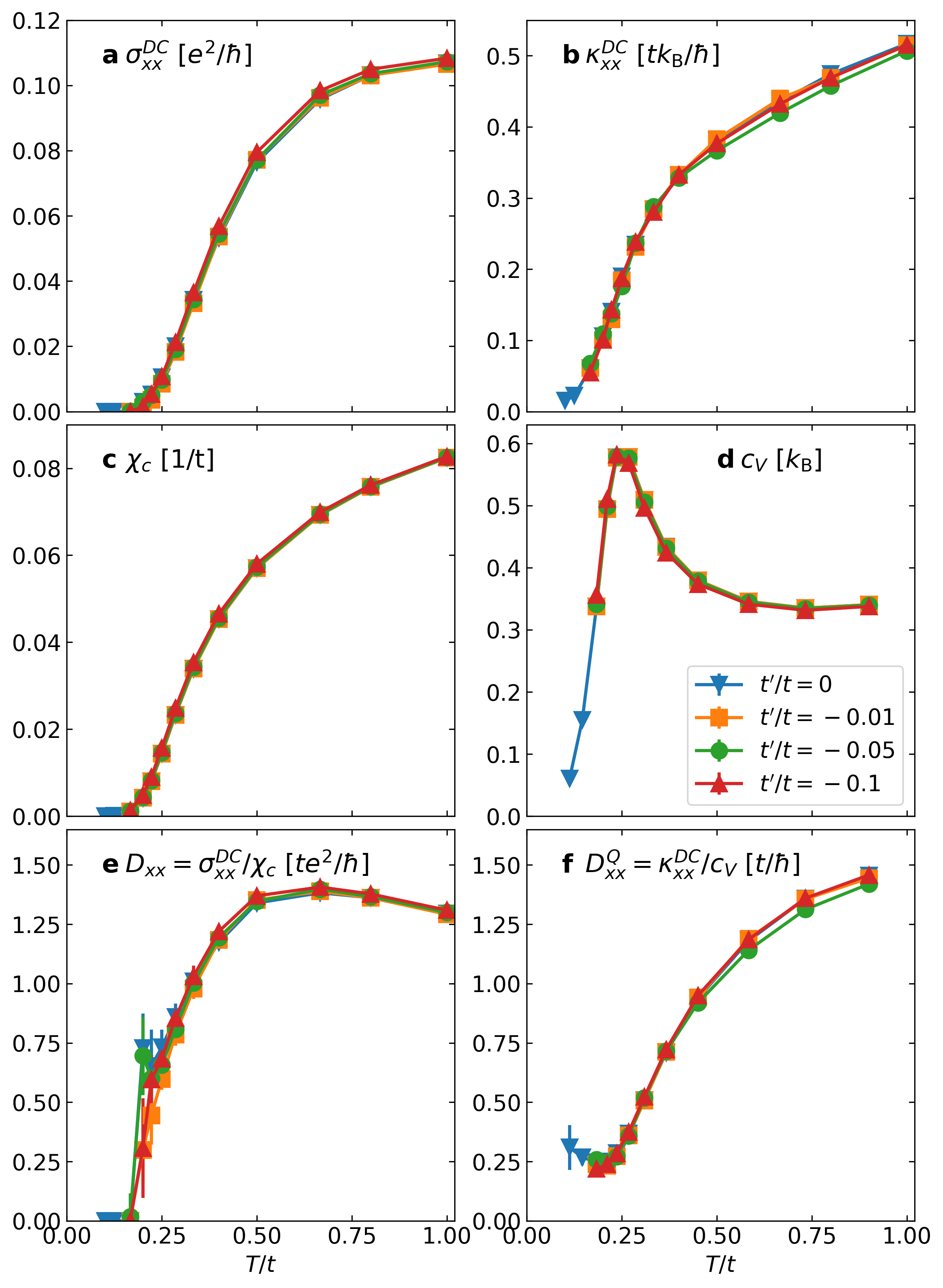}
    \caption[Different trends in longitudinal charge and heat transport]{\textbf{a} Longitudinal DC electrical conductivity $\sigma^{\mathrm{DC}}_{xx}$, \textbf{b} longitudinal DC thermal conductivity $\kappa^{\mathrm{DC}}_{xx}$, \textbf{c} charge compressibility $\chi_c$, \textbf{d} specific heat $c_V$, \textbf{e} charge diffusivity $D^{xx}$, and \textbf{f}thermal diffusivity $D^Q_{xx}$ in the Hubbard-Hofstadter model with $U/t = 6$ at half-filling $\langle n \rangle = 1$ and magnetic field strength $\Phi/\Phi_0 = 4/64$. All panels share the same legend.}
    \label{fig:DC-T-dep-var-tp}
\end{figure}

To further examine the divergent behavior of charge and heat transport, in \cref{fig:DC-T-dep-var-tp} we show the DC ($\omega\rightarrow 0$) conductivities, charge compressibility $\chi_c$, specific heat $c_V$, and electrical and thermal diffusivities $D$ and $D^{Q}$, extracted using 
\begin{equation}
    D = \frac{\sigma}{\chi_c}, \quad D^Q = \frac{\kappa}{c_V}.
\end{equation} 
By comparing \cref{fig:DC-T-dep-var-tp}\textbf{e}-\textbf{f}, we see that at the lowest temperatures we access, the charge diffusivity approaches zero while the thermal diffusivity exhibits an upturn. This further confirms that at $T/t \lesssim 0.2$, we are dealing with an electrical insulator and thermal conductor, with the longitudinal thermal conduction well-understood in terms of magnons~\cite{Wang2022}.

\begin{figure*}[htbp]
    \centering
    \includegraphics[width=0.95\linewidth]{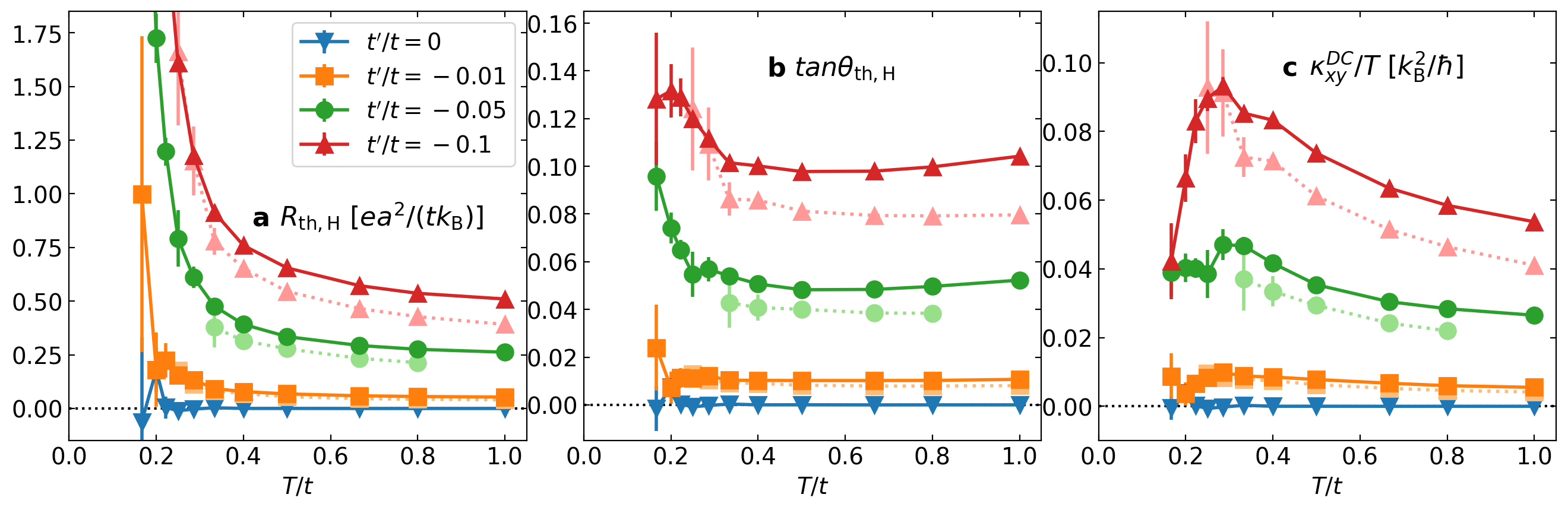}
    \caption[DC thermal Hall coefficient, thermal Hall angle, and thermal Hall conductivity]{\textbf{a} DC thermal Hall coefficient, \textbf{b} thermal Hall angle, and \textbf{c} thermal Hall conductivity. Solid lines denote results obtained by the proxy method, while dotted lines denote results obtained by the subtraction method (See SM~\cite{SuppMats} for detailed methods). Hubbard $U/t = 6$ and field strength $\Phi/\Phi_0 = 4/64$. All panels share the same legend. 
    }
    \label{fig:headline}
\end{figure*}

Next, we use two different methods (see SM~\cite{SuppMats}, Section S4) to obtain the DC thermal Hall conductivity $\kappa^{\mathrm{DC}}_{xy}$, thermal Hall angle $\theta_{\text{th,H}}$, and thermal Hall coefficient $R_{\text{th,H}}$, and show the results in \cref{fig:headline}. The two methods produce qualitatively similar results: while the thermal Hall response is zero at all temperatures when $t'=0$ (within error bars), the thermal Hall response is generically positive and nonzero, with a magnitude that increases with increasing $\abs{t'}$. The high-temperature behavior may be attributed to thermally excited hole-like charge carriers inherited from the underlying tight-binding band structure, which orbitally couple to the magnetic field. However, the low temperature ($T\lesssim J$) thermal Hall response, which we have argued is due to heat transport by magnons, evidently violates the no-go result~\cite{Katsura2010}. We have verified setting Hubbard interaction $U/t = 8$ produces qualitatively similar results as the $U/t=6$ case shown here, and finite-size effects are minimal (see SM~\cite{SuppMats} for additional plots). These checks gives us confidence that the observed nonzero thermal Hall effect at low temperatures are neither remnant signatures of charge fluctuations nor finite-size artifacts.

\textit{Discussion.}--- How do we reconcile the apparent contradiction between our numerical results in \cref{fig:headline} and the no-go result~\cite{Katsura2010}? A careful reading reveals that the no-go result is a restricted statement about linear spin-wave theory, which does not account for effects due to magnon-magnon scattering. 

Performing a strong-coupling expansion on the Hubbard-Hofstadter model shows that the low-energy effective spin Hamiltonian to order $\mathcal{O}(t^3/U^2)$ includes a scalar spin chirality term~\cite{Sen1995,Motrunich2006} 
\begin{equation}
    H_{\mathrm{eff}} = J_1 \sum_{\langle ij\rangle} \vec{S}_i  \cdot \vec{S}_j + J_2 \sum_{\langle \langle ij \rangle \rangle} \vec{S}_i  \cdot \vec{S}_j 
    + J_\chi \sum_{\triangle_{ijk}} \vec{S}_i \cdot (\vec{S}_j\times \vec{S}_k), \label{eq:spin-J1-J2-Jchi}
\end{equation}
where
\begin{equation}
    J_1 = \frac{4t^2}{U}, \ J_2 = \frac{4t'^2}{U}, \ J_\chi = \frac{24t^2t'}{U^2}\sin(\pi\Phi/\Phi_0),
\end{equation}
and $\triangle_{ijk}$ denotes a triangular plaquette with lattice sites $i,j,k$ in anticlockwise order. At the quadratic level, $J_\chi$ does not endow the magnon bands with nontrivial Berry curvature required for thermal Hall transport~\cite{Matsumoto2014}, as it vanishes both in linear spin-wave theory~\cite{Katsura2010} and at the mean-field level~\cite{SuppMats}. However, it was recently discovered that magnon-magnon scattering, which does not depend on topological band theory, may give rise to a finite thermal Hall coefficient~\cite{Chatzichrysafis2024}. 

As a minimal example, we illustrate this mechanism using semi-classical Boltzmann transport, in which the thermal Hall conductivity is related to the rate of magnon mode collisions (or collision kernel) by~\cite{Chatzichrysafis2024}
\begin{equation}
    \kappa_{xy} = \frac{1}{2 k_B T^2 V}\sum_{\mathbf{k},\mathbf{k}'}
    (\textbf{v}_\textbf{k}\times\textbf{v}_{\textbf{k}'})_z \varepsilon_\textbf{k}\varepsilon_{\textbf{k}'}\tau_\textbf{k}\tau_{\textbf{k}'}G_\textbf{k}G_{\textbf{k}'}\mathcal{A}_{\textbf{k}\textbf{k}'},
\end{equation}
where $\varepsilon_\textbf{k}$ is the dispersion, $\textbf{v}_\textbf{k}=\partial\varepsilon_\textbf{k}/\partial\textbf{k}$ is the magnon group velocity, $\tau_\textbf{k}$ is the magnon lifetime, and $G_\textbf{k}=\sqrt{\overline{N}_\textbf{k}(\overline{N}_\textbf{k}+1)}$, where $\overline{N}_\textbf{k}$ is the Bose-Einstein distribution. Here, $\mathcal{A}_{\textbf{k}\textbf{k}'}=(\mathcal{O}_{\textbf{k}{\textbf k}'}-\mathcal{O}_{\textbf{k}'{\textbf k}})/2$ is the antisymmetric part of the collision kernel, where $\mathcal{O}_{\textbf{kk}'}$ is the total off-diagonal scattering rate, composed of $\mathcal{O_{\mathbf{kk'}}^{++}},\mathcal{O_{\mathbf{kk'}}^{--}},\mathcal{O_{\mathbf{kk'}}^{+-}}$, and $\mathcal{O_{\mathbf{kk'}}^{-+}}$. The $\pm$ superscripts describe scattering events where modes with the corresponding momentum subscripts are created (+) or destroyed (-). These scattering rates may be computed using Fermi's Golden Rule~\cite{Chatzichrysafis2024}. 

It can be shown that in order for $\mathcal{A}_{\textbf{k}\textbf{k}'}$, and therefore $\kappa_{xy}$, to be finite, the microscopic detailed balance relations
\begin{equation}
    \frac{\mathcal{O_{\mathbf{kk'}}^{++}}}{\mathcal{O_{\mathbf{kk'}}^{--}}}=e^{-\beta\varepsilon_{\mathbf{k'}}}\qquad\frac{\mathcal{O_{\mathbf{kk'}}^{+-}}}{\mathcal{O_{\mathbf{kk'}}^{-+}}}=e^{\beta\varepsilon_{\mathbf{k'}}}\label{eq:db}
\end{equation}
must be violated~\cite{Mangeolle2022,Mangeolle2022PRB,Chatzichrysafis2024}. If we consider the $J_1$-$J_2$-$J_\chi$ model, we find that only the collisions mediated by the $J_\chi$ term violate Eq. \eqref{eq:db}. In other words, one cannot generate a finite $\kappa_{xy}$ from magnon-magnon scattering without both $t' \neq 0$ and $\Phi\neq 0$ on the square lattice, which is consistent with our numerical results and symmetry argument. Moreover, the leading order contribution is an interference process between a first and second order scattering event, implying $\kappa_{xy}\propto J_\chi J_1^2\sim t'\sin(\pi \Phi / \Phi_0)$. The linear magnetic field dependence (when $\Phi \ll \Phi_0$) and linear $t'$ dependence of $\kappa_{xy}$ are consistent with our numerical results, as shown in SM~\cite{SuppMats}, Figure S11, and  Fig. \ref{fig:headline}, respectively.

We emphasize that the magnon-magnon scattering picture from the downfolded spin-Hamiltonian is only one mechanism for generating a finite thermal Hall. In contrast, our DQMC results are valid independently of any specific thermal current carriers or the assumptions of Boltzmann scattering, and is a more general demonstration of the violation of the no-go theorem. 

While it is tempting to directly compare the magnitude of the thermal Hall conductivity $\kappa_{xy}/T$ we obtain to experimental values~\cite{Grissonnanche2019}, we emphasize that when converted to units appropriate to cuprate materials, e.g. $t/k_B\sim 4000 \si{\kelvin}$ and $a = 3.8$\r{A}, our lowest temperature corresponds to $T\sim \SI{700}{\kelvin}$, and our lowest magnetic field strength corresponds to $B \sim \SI{400}{\tesla}$. It is not at all straightforward to extrapolate our results to experimentally reasonable temperatures $T\lesssim \SI{100}{\kelvin}$ and field strengths $B \sim \SI{10}{\tesla}$, so we won't attempt to do so. The main focus of this work is a demonstration of the intrinsic mechanism of the thermal Hall effect, rather than a quantitative prediction of the thermal Hall signal in the cuprates. Our work primarily serves as a ``proof of principle'': we establish that the $t$-$t'$-$U$ Hubbard model on the square lattice exhibits a nonzero thermal Hall effect under an applied magnetic field. Therefore, in analyzing experimental data, one should not naively ignore potential magnon contributions to the thermal Hall effect based on the no-go theorem~\cite{Katsura2010}, which has a much narrower regime of validity than commonly interpreted.

\textit{Data Availability.}--- Aggregated numerical data and analysis routines required to reproduce the figures can be found at \url{10.5281/zenodo.13799597}. Raw simulation data that support the findings of this study are stored on the Sherlock cluster at Stanford University and are available from the corresponding author upon reasonable request.

\textit{Code Availability.}--- The most up-to-date version of our thermal transport DQMC simulation code can be accessed at \url{https://github.com/katherineding/dqmc-dev}.

\textit{Acknowledgements.}--- We are grateful for helpful discussions with Steve Kivelson, Alexander Balatsky, Anjishnu Bose, Junkai Dong, Nishchhal Verma, Vladimir Calvera, Alexander Mook, and Julian May-Mann. 

This work was supported by the U.S. Department of Energy (DOE), Office of Basic Energy Sciences,
Division of Materials Sciences and Engineering. 
Computational work was performed on the Sherlock cluster at Stanford University and on resources of the National Energy Research Scientific Computing Center (NERSC), a Department of Energy Office of Science User Facility, using NERSC award BES-ERCAP0027200. E.Z.Z. and Y.B.K are supported by the Natural Science and Engineering Research Council (NSERC) of Canada and the Center for Quantum Materials at the University of Toronto. E.Z.Z. was further supported by the Michael Smith Foreign Study Scholarship. T.C. is supported by a University of California Presidential Postdoctoral Fellowship and acknowledges support from the Gordon and Betty Moore Foundation through Grant No. GBMF8690 to UC Santa Barbara.
\bibliography{main}

\end{document}


\title{\bf Supplementary Material to ``Intrinsic Thermal Hall Effect in Mott Insulators''}
\date{\vspace{-1cm}}
\maketitle

\section{Simulation Parameters}
\label{sec:params}
\begin{figure}[htbp]
    \centering
    \includegraphics[width=0.8\linewidth]{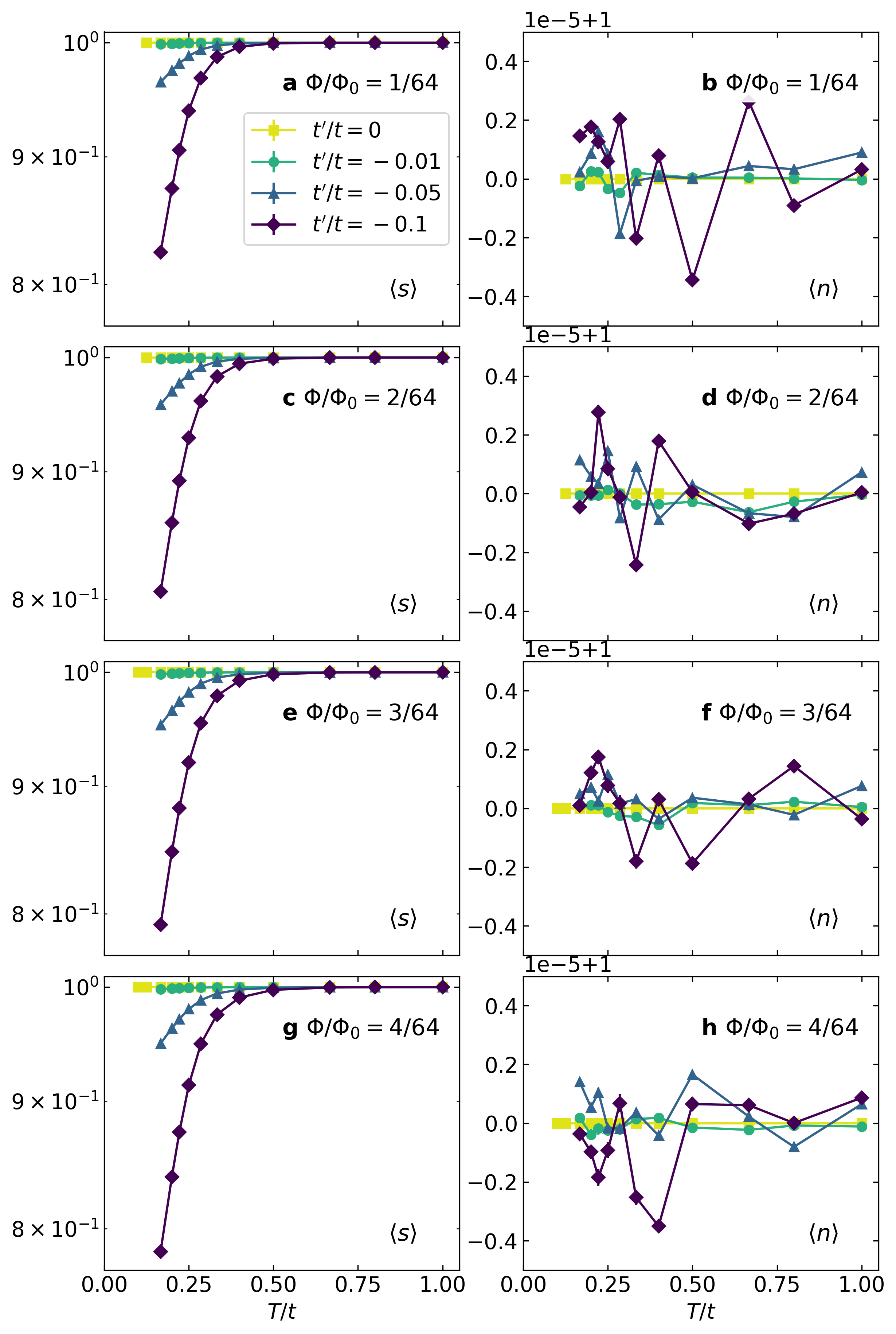}
    \caption[Fermion sign and average particle density]{Fermion sign and average particle density after chemical potential tuning. Hubbard $U/t = 6$, target half filling $\expval{n}=1$. Errorbars (smaller than the size of the data points) denote one standard deviation of the mean, determined by jackknife resampling. }
    \label{fig:sign-density-U6}
\end{figure}

Determinant quantum Monte Carlo (DQMC) data shown in main text figures are obtained from simulations performed using $2\times 10^4$ to $5\times 10^4$ warm-up sweeps  and $3\times 10^5$ to $5 \times 10^6$ measurement sweeps through the auxillary field. We run $120$ to $500$ independently seeded Markov chains for each set of parameters. For all parameter values, the imaginary time discretization interval $\Delta\tau \leq 0.05/t$, and the number of imaginary time slices $L = \beta/\Delta \tau \geq 10$. Such a small imaginary time discretization interval is chosen in order to reduce effects from Trotter error. The chemical potential is fine tuned so that particle density satisfies $\abs{\langle n \rangle  - 1} <  4\times 10^{-5}$, as shown in \cref{fig:sign-density-U6}.

In all simulations, multiple equal-time measurements are taken in each full measurement sweep through the auxillary field, while unequal-time measurements are taken every few full measurement sweeps. Specifically, each Markov chain with $M$ measurement sweeps collects $M L /5$ equal-time measurements, and $M/2$ unequal-time measurements. The mean and standard error of equal-time observables and the finite-Matsubara-frequency proxy are estimated via jackknife resampling of independent Markov chains. The mean and standard error of MaxEnt results are estimated via bootstrap resampling of independent Markov chains using 100 bootstrap samples.
In MaxEnt fits, we always use flat model functions and choose hyper-parameter $\alpha$ using the ``BT'' method~\cite{Bergeron2016}. For the subtraction method described in \cref{subsec:subtract}, the same bootstrap resamples are used for both the composite object and the longitudinal response. 

\section{Electrical Current Operator}
\label{sec:electric-current}

Charge current is carried by particle current: $\vec{J} = (-e) \vec{J}_N$. The particle current operator is easy to find, e.g. see ref.~\cite{Mahan2013}, where we define a polarization operator $\vec{P} = \sum_j \vec{R}_j n_j$ and find $\vec{J}_N = \dfrac{i}{\hbar}[H, \vec{P}]$.
Written out explicitly, the total charge current operator is 
\begin{align}
\vec{J} &= \frac{i(-e)}{\hbar}\sum_{ij\sigma} t_{ij}\exp\left[i\varphi_{ij}\right](\vec{R}_i - \vec{R}_j)c_{i\sigma}^\dagger c_{j\sigma} \label{eq:current-vector}\\
&= \frac{ie}{2\hbar} \sum_{ij\sigma} t_{ij} (\vec{R}_j -\vec{R}_i) \left[ \exp[i\varphi_{ij}] c_{i\sigma}^\dagger c_{j\sigma} - \exp[i\varphi_{ji}] c_{j\sigma}^\dagger c_{i\sigma}\right],
\end{align}
which componentwise becomes
\begin{align}
J_x = \frac{i(-e)}{\hbar} \sum_{ij\sigma}t_{ij}\exp\left[i\varphi_{ij}\right]c_{i\sigma}^\dagger c_{j\sigma}(x_i - x_j) \label{eq:current-x},\\
J_y = \frac{i(-e)}{\hbar} \sum_{ij\sigma}t_{ij}\exp\left[i\varphi_{ij}\right]c_{i\sigma}^\dagger c_{j\sigma}(y_i-y_j) \label{eq:current-y}.
\end{align}

Current operators are Hermitian, even in presence of nonzero magnetic field. So we can always write
\begin{equation}
J_x^\dagger = J_x, \qquad J_y^\dagger = J_y.
\end{equation}

\section{Heat Current Operator}
\label{sec:heat-current}
Energy currents flow whenever heat is generated or dissipated \textit{non-uniformly} in the solid. (Total) energy current $\vec{J}_E$ has units [energy $\cdot$ velocity] and obeys continuity equation~\cite{Mahan2013} (Here $H$ should not include the chemical potential term):
\begin{equation*}
\frac{\partial}{\partial  t} H + \nabla \cdot \vec{J}_E = 0.
\end{equation*}
Completely analogous to how we defined a polarization operator to find the expression for the particle current operator, we formally introduce an operator $\vec{R}_E$,
\begin{equation}
\vec{R}_E = \frac{1}{2}\int d\vec{r} \left[\vec{r} \mathcal{H}(\vec{r}) + \mathcal{H}(\vec{r})\vec{r} \right],
\end{equation}
where $\mathcal{H}(\vec{r})$ is the hamiltonian/energy density, which allows us to find the energy current via
\begin{equation}
\frac{d\vec{R}_E}{dt} = \frac{i}{\hbar}\left[H,\vec{R}_E\right] = \vec{J}_E.
\end{equation}
What is $\vec{R}_E$ in the Hubbard model? Using
\begin{align}
H &= -\sum_{ ij \sigma} t_{ij}\exp[\varphi_{ij}]c_{i \sigma}^\dagger c_{j \sigma} + U\sum_{i}\left(c^\dagger_{i \uparrow}c_{i\uparrow} - 1/2 \right)\left(c^\dagger_{i \downarrow}c_{i\downarrow} - 1/2\right), \\
h_i &= -\frac{1}{2}\sum_{j\sigma} t_{ij}  \left(\exp[\varphi_{ij}]c^\dagger_{i\sigma} c_{j\sigma}+\exp[\varphi_{ji}]c^\dagger_{j\sigma} c_{i\sigma}\right) + U \left(c^\dagger_{i\uparrow} c_{i\uparrow}  - \frac{1}{2}\right)\left(c^\dagger_{i\downarrow} c_{i\downarrow} - \frac{1}{2}\right), \label{eq:local-energy-hi}
\end{align}
we find 
\begin{equation}
    \vec{R}_E = \sum_i \vec{R}_i h_i = -\frac{1}{2}\sum_{ij\sigma} \vec{R}_i t_{ij}  \left(\exp[\varphi_{ij}]c^\dagger_{i\sigma} c_{j\sigma}+\exp[\varphi_{ji}]c^\dagger_{j\sigma} c_{i\sigma}\right) + U \sum_i \vec{R}_i\left(c^\dagger_{i\uparrow} c_{i\uparrow}  - \frac{1}{2}\right)\left(c^\dagger_{i\downarrow} c_{i\downarrow} - \frac{1}{2}\right) .
\end{equation}
Note that we wrote $h_i$ in the form \cref{eq:local-energy-hi}  because we desire that it be explicitly hermitian $h_i = h_i^\dagger$.
Using the above allows us to get $\vec{J}_E$:
\begin{align}
\vec{J}_E &= \frac{i}{\hbar} \left[H - \mu N, \vec{R}_E\right] = \nonumber \\
&= \frac{i}{\hbar} \left[-\sum_{ ij \sigma} t_{ij}\exp[\varphi_{ij}]c_{i \sigma}^\dagger c_{j \sigma} - \mu \sum_{i\sigma} c_{i\sigma}^\dagger c_{i\sigma} + U\sum_{i}\left(c^\dagger_{i \uparrow}c_{i\uparrow} - 1/2 \right)\left(c^\dagger_{i \downarrow}c_{i\downarrow} - 1/2\right), \right. \nonumber \\
& \qquad \quad \left.-\frac{1}{2}\sum_{ij\sigma} \vec{R}_i t_{ij}  \left(\exp[\varphi_{ij}]c^\dagger_{i\sigma} c_{j\sigma}+\exp[\varphi_{ji}]c^\dagger_{j\sigma} c_{i\sigma}\right) + U \sum_i \vec{R}_i\left(c^\dagger_{i\uparrow} c_{i\uparrow}  - \frac{1}{2}\right)\left(c^\dagger_{i\downarrow} c_{i\downarrow} - \frac{1}{2}\right) \right] \nonumber \\
&=\frac{i}{\hbar} \left[-\sum_{ ij \sigma} t_{ij}\exp[\varphi_{ij}]c_{i \sigma}^\dagger c_{j \sigma} - \left(\mu + \frac{1}{2}U\right) \sum_{i\sigma} n_{i\sigma} + U\sum_{i}n_{i \uparrow} n_{i\downarrow} , \right. \nonumber\\
& \qquad \quad \left.-\frac{1}{2}\sum_{ij\sigma} \vec{R}_i t_{ij}  \left(\exp[\varphi_{ij}]c^\dagger_{i\sigma} c_{j\sigma}+\exp[\varphi_{ji}]c^\dagger_{j\sigma} c_{i\sigma}\right)-  \frac{1}{2}U \sum_{i\sigma} \vec
{R}_i n_{i\sigma} + U \sum_i \vec{R}_i n_{i \uparrow} n_{i\downarrow} \right] .
\end{align}
There are 3 terms in Hamiltonian, 3 terms in $\vec{R}_E$. We label the 3 terms in the Hamiltonian ``K,'' ``N,'' and ``U,'' and the 3 terms in $\vec{R}_E$ ``RK,'' ``RN,'' and ``RU.'' The commutator has 9 terms in total; 4 of them involve only commutation between number operators (namely the ``N''-``RN,'' ``N''-``RU,'' ``U''-``RN,'' ``U''-``RU'' combinations) are identically zero. We compute the rest manually.

We make use of fermion commutator relations
\begin{align*}
\left[ c_{i\sigma} ^\dagger, n_{k\sigma'}\right] = - \delta_{ik} \delta_{\sigma \sigma'} c^\dagger_{k\sigma'} , \qquad\qquad
\left[ c_{i\sigma}, n_{k\sigma'}\right] = \delta_{ik} \delta_{\sigma \sigma'} c_{k\sigma'}, 
\end{align*}
and the following derived identities:
\begin{align}
\left[c_{i\sigma}^\dagger c_{j\sigma},n_{k\sigma'}\right] &= \left( c_{i\sigma}^\dagger c_{k\sigma'}\delta_{jk} -c_{k\sigma'}^\dagger c_{j\sigma}\delta_{ik} \right)\delta_{\sigma\sigma'} \label{commute-1}, \\
\left[c_{i\sigma}^\dagger c_{j\sigma}, n_{k\uparrow}n_{k\downarrow}\right] &= 
n_{k\downarrow}\delta_{\sigma \uparrow} \left( c_{i\sigma}^\dagger c_{k\uparrow} \delta_{jk} 
- c_{k\uparrow}^\dagger c_{j\sigma} \delta_{ik} \right)
+ n_{k\uparrow}\delta_{\sigma\downarrow}\left(c_{i\sigma}^\dagger c_{k\downarrow} \delta_{jk} - c_{k\downarrow}^\dagger c_{j\sigma}\delta_{ik}\right) \label{commute-2}, \\
\left[c_{i\sigma}^\dagger c_{j\sigma},c_{m\sigma'}^\dagger c_{n\sigma'}\right] &= \left(c_{i\sigma}^\dagger c_{n\sigma'}\delta_{mj} - c_{m\sigma'}^\dagger c_{j\sigma} \delta_{ni}\right)\delta_{\sigma\sigma'}\label{commute-3}.
\end{align}
\cref{commute-1} tells us that the ``N''-``RK'' cross term is
\begin{equation}
\left[\left(\mu + \frac{1}{2}U\right) \sum_{k\sigma'} n_{k\sigma'},-\frac{1}{2}\sum_{ij\sigma} \vec{R}_i t_{ij}  \left(\exp[\varphi_{ij}]c^\dagger_{i\sigma} c_{j\sigma}+\exp[\varphi_{ji}]c^\dagger_{j\sigma} c_{i\sigma}\right)\right] =0,
\end{equation}
and the ``K''-``RN'' cross term is
\begin{equation}
\left[ -\sum_{ ij \sigma} t_{ij}\exp[\varphi_{ij}]c_{i \sigma}^\dagger c_{j \sigma} ,-\frac{1}{2}U \sum_{k\sigma} \vec
{R}_k n_{k\sigma}\right] = 
\frac{1}{2}U \sum_{ij\sigma} t_{ij} \exp[i\varphi_{ij}]\left(\vec{R}_j - \vec{R}_i\right)c_{i \sigma}^\dagger c_{j \sigma} .
\end{equation}

The ``K''-``RU'' and ``U''-``RK'' kinetic-double occupancy cross terms are a little more involved. We compute using \cref{commute-2}
\begin{align*}
&\left[-\sum_{ ij \sigma} t_{ij}\exp[\varphi_{ij}]c_{i \sigma}^\dagger c_{j \sigma},U \sum_k \vec{R}_k n_{k \uparrow} n_{k\downarrow} \right] + 
\left[U\sum_{k}n_{k \uparrow} n_{k\downarrow},-\frac{1}{2}\sum_{ij\sigma} \vec{R}_i t_{ij}  \left(\exp[\varphi_{ij}]c^\dagger_{i\sigma} c_{j\sigma}+\exp[\varphi_{ji}]c^\dagger_{j\sigma} c_{i\sigma}\right) \right]\\
&=-U\sum_{ijk\sigma} t_{ij}\exp[i\varphi_{ij}] \vec{R}_k\left[c_{i \sigma}^\dagger c_{j \sigma}, n_{k \uparrow} n_{k\downarrow} \right] -\frac{1}{2}U\sum_{ijk\sigma} t_{ij}\exp[i\varphi_{ij}] \left(\vec{R}_i + \vec{R}_j\right)\left[ n_{k \uparrow} n_{k\downarrow},c_{i \sigma}^\dagger c_{j \sigma}\right]\\
&=-U\sum_{ijk\sigma} t_{ij}\exp[i\varphi_{ij}] \left[\vec{R}_k-\frac{1}{2}\left(\vec{R}_i + \vec{R}_j\right)\right]\left[c_{i \sigma}^\dagger c_{j \sigma}, n_{k \uparrow} n_{k\downarrow} \right] \\
&=-\frac{1}{2}U \sum_{ij} t_{ij} \exp[i\varphi_{ij}]\left(\vec{R}_j - \vec{R}_i\right) \left[(n_{i\downarrow} + n_{j\downarrow}) c_{i\uparrow}c_{j\uparrow} + (n_{i\uparrow} + n_{j\uparrow}) c_{i\downarrow}c_{j\downarrow}\right]\\
&=-\frac{1}{2}U \sum_{ij\sigma} t_{ij} \exp[i\varphi_{ij}]\left(\vec{R}_j - \vec{R}_i\right) (n_{i\sigma} + n_{j\sigma}) c_{i\bar{\sigma}}^\dagger c_{j\bar{\sigma}}.
\end{align*}

Finally, there is the ``K''-``RK'' term. Using \cref{commute-3} we have:
\begin{align}
&\left[-\sum_{ ij \sigma} t_{ij}\exp[i\varphi_{ij}] c_{i \sigma}^\dagger c_{j \sigma},-\sum_{mn\sigma'} \vec{R}_m t_{mn}\exp[i\varphi_{mn}] c^\dagger_{m\sigma} c_{n\sigma}\right] \nonumber\\
&=\sum_{ijmn \sigma}  t_{ij}\exp[i\varphi_{ij}]\vec{R}_m t_{mn}\exp[i\varphi_{mn}] \left(c_{i\sigma}^\dagger c_{n\sigma}\delta_{mj} - c_{m\sigma}^\dagger c_{j\sigma} \delta_{ni}\right) \nonumber\\
&= \sum_{ij n \sigma}  t_{ij}\exp[i\varphi_{ij}]\vec{R}_j t_{jn}\exp[i\varphi_{jn}] c_{i\sigma}^\dagger c_{n\sigma} - \sum_{ij m \sigma}  t_{ij}\exp[i\varphi_{ij}]\vec{R}_m t_{mi}\exp[i\varphi_{mi}] c_{m\sigma}^\dagger c_{j\sigma} \nonumber\\
&= \sum_{ijk\sigma} t_{ij} t_{jk} \exp[i\varphi_{ij}]\exp[i\varphi_{jk}] \left(\vec{R}_j - \vec{R}_i\right) c_{i\sigma}^\dagger c_{k\sigma} \label{kk-part1},
\end{align}
and
\begin{align}
&\left[-\sum_{ ij \sigma} t_{ij}\exp[i\varphi_{ij}] c_{i \sigma}^\dagger c_{j \sigma},-\sum_{mn\sigma'} \vec{R}_m t_{mn}\exp[i\varphi_{nm}] c^\dagger_{n\sigma} c_{m\sigma}\right] \nonumber\\
&=\sum_{ijmn \sigma}  t_{ij}\exp[i\varphi_{ij}]\vec{R}_m t_{mn}\exp[i\varphi_{nm}] \left(c_{i\sigma}^\dagger c_{m\sigma}\delta_{nj} - c_{n\sigma}^\dagger c_{j\sigma} \delta_{mi}\right) \nonumber\\
&= \sum_{ij m \sigma}  t_{ij}\exp[i\varphi_{ij}]\vec{R}_m t_{mj}\exp[i\varphi_{jm}] c_{i\sigma}^\dagger c_{m\sigma} - \sum_{ij n \sigma}  t_{ij}\exp[i\varphi_{ij}]\vec{R}_i t_{in}\exp[i\varphi_{ni}] c_{n\sigma}^\dagger c_{j\sigma} \nonumber\\
&= \sum_{ijk\sigma} t_{ij} t_{jk} \exp[i\varphi_{ij}]\exp[i\varphi_{jk}] \left(\vec{R}_k - \vec{R}_j\right) c_{i\sigma}^\dagger c_{k\sigma} \label{kk-part2}.
\end{align}

Putting \cref{kk-part1,kk-part2} together, we have
\begin{align*}
&\left[-\sum_{ ij \sigma} t_{ij}\exp[\varphi_{ij}]c_{i \sigma}^\dagger c_{j \sigma},-\frac{1}{2}\sum_{mn\sigma} \vec{R}_m t_{mn}  \left(\exp[\varphi_{mn}]c^\dagger_{m\sigma} c_{n\sigma}+\exp[\varphi_{nm}]c^\dagger_{n\sigma} c_{m\sigma}\right)\right]\\
&=\frac{1}{2}\sum_{ijk\sigma} t_{ij} t_{jk} \exp[i\varphi_{ij}]\exp[i\varphi_{jk}] \left(\vec{R}_k - \vec{R}_i\right) c_{i\sigma}^\dagger c_{k\sigma}.
\end{align*}

Summing all terms together, we have
\begin{align*}
\vec{J}_E &= \frac{i}{\hbar}\left\{\frac{1}{2}\sum_{ijk\sigma} t_{ij} t_{jk} \exp[i\varphi_{ij}]\exp[i\varphi_{jk}] \left(\vec{R}_k - \vec{R}_i\right) c_{i\sigma}^\dagger c_{k\sigma}\right.\\
&\qquad \quad-\frac{1}{2}U \sum_{ij\sigma} t_{ij} \exp[i\varphi_{ij}]\left(\vec{R}_j - \vec{R}_i\right) (n_{i\sigma} + n_{j\sigma}) c^\dagger_{i\bar{\sigma}}c_{j\bar{\sigma}} \\
&\left.\qquad \quad+\frac{1}{2}U \sum_{ij\sigma} t_{ij} \exp[i\varphi_{ij}]\left(\vec{R}_j - \vec{R}_i\right)c_{i \sigma}^\dagger c_{j \sigma} \right\}\\
&= \frac{i}{\hbar}\left\{\frac{1}{4}\sum_{ijk\sigma} t_{ij} t_{jk}  \left(\vec{R}_k - \vec{R}_i\right) \left[\exp[i\varphi_{ij}]\exp[i\varphi_{jk}]c_{i\sigma}^\dagger c_{k\sigma} -\exp[i\varphi_{ji}]\exp[i\varphi_{kj}]c_{k\sigma}^\dagger c_{i\sigma}\right]\right.\\
&\quad \qquad-\frac{1}{4}U \sum_{ij\sigma} t_{ij} \left(\vec{R}_j - \vec{R}_i\right) (n_{i\sigma} + n_{j\sigma})\left[ \exp[i\varphi_{ij}]c^\dagger_{i\bar{\sigma}}c_{j\bar{\sigma}}-\exp[i\varphi_{ji}]c^\dagger_{j\bar{\sigma}}c_{i\bar{\sigma}}\right]\\
&\left.\quad  \qquad+\frac{1}{4}U \sum_{ij\sigma} t_{ij} \left(\vec{R}_j - \vec{R}_i\right)\left[\exp[i\varphi_{ij}]c_{i \sigma}^\dagger c_{j \sigma}-\exp[i\varphi_{ji}]c_{j \sigma}^\dagger c_{i\sigma}\right]\right\}.
\end{align*}

But energy current is not the current which describes thermal conductivity or thermoelectric power. The (total) heat current is defined using (total) energy current and (total) particle current as 
\begin{equation*}
\vec{J}_Q = \vec{J}_E - \mu \vec{J}_N .
\end{equation*}

Particle current $\vec{J}_N$ is (see \cref{sec:electric-current})
\begin{align*}
\vec{J}_N &= -\frac{i}{\hbar}\sum_{ij\sigma} t_{ij} \exp[i\varphi_{ij}] \left(\vec{R}_j -\vec{R}_i\right) c_{i\sigma}^\dagger c_{j\sigma} \\
&= -\frac{i}{2\hbar}\sum_{ij\sigma} t_{ij} \left(\vec{R}_j -\vec{R}_i\right) \left[\exp[i\varphi_{ij}] c_{i\sigma}^\dagger c_{j\sigma}- \exp[i\varphi_{ji}] c_{j\sigma}^\dagger c_{i\sigma}\right],
\end{align*}
so the heat current $\vec{J}_Q$ is
\begin{align}
\vec{J}_Q &= \frac{i}{\hbar}\left\{\frac{1}{4}\sum_{ijk\sigma} t_{ij} t_{jk}  \left(\vec{R}_k - \vec{R}_i\right) \left[\exp[i\varphi_{ij}]\exp[i\varphi_{jk}]c_{i\sigma}^\dagger c_{k\sigma} -\exp[i\varphi_{ji}]\exp[i\varphi_{kj}]c_{k\sigma}^\dagger c_{i\sigma}\right]\right. \nonumber\\
&\quad \qquad-\frac{1}{4}U \sum_{ij\sigma} t_{ij} \left(\vec{R}_j - \vec{R}_i\right) (n_{i\sigma} + n_{j\sigma})\left[ \exp[i\varphi_{ij}]c^\dagger_{i\bar{\sigma}}c_{j\bar{\sigma}}-\exp[i\varphi_{ji}]c^\dagger_{j\bar{\sigma}}c_{i\bar{\sigma}}\right] \nonumber\\
&\left.\quad  \qquad+\frac{1}{4}\left(U +2\mu\right)\sum_{ij\sigma} t_{ij} \left(\vec{R}_j - \vec{R}_i\right)\left[\exp[i\varphi_{ij}]c_{i \sigma}^\dagger c_{j \sigma}-\exp[i\varphi_{ji}]c_{j \sigma}^\dagger c_{i\sigma}\right]\right\} \label{heat-current}.
\end{align}

Heat current operators are Hermitian, even in presence of nonzero magnetic field. So we can always write
\begin{equation}
J_{Q,x}^\dagger = J_{Q,x}, \qquad J_{Q,y}^\dagger = J_{Q,y}.
\end{equation}

\section{Detailed Methods}
\label{eq:detailed-methods}
 
A description of linear response, Kubo formulas, trasnsport theory can be found in standard textbooks \cite{Mahan2013,Marder2010,Shastry2008}, but it's important to keep the sign, normalization, and notational conventions consistent throughout, so we briefly describe this formalism below.

We first note that computing the response to a nonuniform temperature requires some care since the thermal gradient does not directly modify the Hamiltonian as a usual perturbation, but rather the Boltzmann factor $e^{-H/k_BT(\mathbf{r})}$. We follow the formalism introduced by Luttinger~\cite{Luttinger1964}, where we consider an expansion to the temperature as $T(\mathbf{r})=T (1-\psi(\mathbf{r}))$, where $\psi(\mathbf{r})$ is a small deviation from $T$, and is also known as a pseudogravitational potential. Up to terms linear in $\psi$, the Boltzmann factor becomes 
\begin{align}
	e^{-H/k_BT(\mathbf{r})}=e^{-H/k_BT (1-\psi(\mathbf{r}))} \simeq e^{-H(1+\psi)/k_BT}. 
\end{align}
Considering the more general case of a time-dependent perturbation, we may now write $H=H_0+F$, where 
\begin{align}
H_0=\int \text{d}\mathbf{r}\ h_0(\mathbf{r}) \qquad  \text{ and} \qquad F = \int \text{d}\mathbf{r}\ \psi(\mathbf{r},t)h_0(\mathbf{r}). \label{eq:pg}
\end{align}
Next, we consider the thermal-electric linear response equations 
\begin{align}
    \vec{j} = L^{(11)} \vec{E} + L^{(12)} \left[-\nabla \psi\right] \label{eq:onsager-1}\\
    \vec{j}_Q = L^{(21)} \vec{E} + L^{(22)} \left[-\nabla \psi\right] \label{eq:onsager-2}
\end{align}
where $\vec{j}$ is the electric current density, $\vec{j}_Q$ is the heat current density, $T$ is temperature, and $\vec{E}$ is electric field.

Note each $L^{(\mu\nu)}$ is itself a matrix, which, for our two dimensional system, has $x$ and $y$ components. 
The coefficients $L^{(\mu\nu)}_{\alpha\beta}$ are generally complex numbers, where $\mu,\nu$ index current type $1,2$ representing charge and heat current respectively, $\alpha,\beta$ index directions $x,y$,
can be computed via Kubo formulas and expressed in terms of retarded current-current operators \footnote{The actual conductivity is the sum of a pole at $\omega=0$ and a regular part. see e.g. \cite{Rigol2008} and references therein. Usually we ignore the pole because its weight goes to zero in the thermodynamic limit, unless the system is a perfect conductor or is a superconductor.}
\begin{equation}
L^{(\mu\nu)}_{\alpha\beta} =  \frac{1}{\hbar\omega V}\int_{-\infty}^\infty dt\, \theta(t) \left\langle \left[J_{\mu,\alpha}^\dagger (t),J_{\nu,\beta} (0)\right]\right\rangle_0 e^{i\omega t} = \frac{i}{\omega}\chi^{R}_{\mu\nu,\alpha\beta}(\omega) \label{eq:kubo},
\end{equation}
where the retarded correlators in real frequency and real time are 
\begin{align}
    \chi^{R}_{\mu\nu,\alpha\beta}(\omega) &= \frac{-i}{\hbar V}\int_{-\infty}^{\infty} dt\, \theta(t) \left\langle \left[J_{\alpha}^\dagger(t), J_{\beta}(0)\right] \right \rangle_0 e^{i\omega t} =\int_{-\infty}^{\infty} dt\, \chi^R_{\mu\nu,\alpha\beta} (t) e^{i\omega t} \label{eq:retard-real-freq},\\
    \chi^R_{\mu\nu,\alpha\beta} (t) &= \frac{-i}{\hbar V}\theta(t) \left\langle \left[J_{\alpha}^\dagger(t), J_{\beta}(0)\right] \right \rangle_0 = \frac{1}{2\pi}\int_{-\infty}^{\infty} d\omega\, \chi^R_{\mu\nu,\alpha\beta} (\omega) e^{-i\omega t} \label{eq:retard-real-time},
\end{align}
and $\vec{J} = \vec{j}(\vec{q} = 0)$ is the total electrical current operator,  $\vec{J}_Q = \vec{j}_q(\vec{q} = 0)$ is the total heat current operator. Operators evolve in time according to the interaction representation, $\expval{}_0$ denotes taking expectation value in the unperturbed thermodynamic ensemble, and $V = Na^2$ is system volume. Derivations and explicit forms of total current operators are shown in \cref{sec:electric-current,sec:heat-current}. 

As $\vec{J}$ has units of $eta/\hbar$, $L^{(11)}$ and $\sigma$ have units of $e^2/\hbar$, and $R_\mathrm{H}$ has units of $a^2/e$. 
As $\vec{J}_Q$ has units of $t^2 a/\hbar$, $L^{(22)}$ has units of $t^2/\hbar$, $\kappa$ has units of $tk_B/\hbar$, $\kappa/T$ has units of $k_B^2/\hbar$, and $R_{\mathrm{th,H}}$ has units of $ea^2/(tk_B)$.

The retarded current-current correlators defined in \cref{eq:retard-real-freq,eq:retard-real-time} can be written in Lehmann/spectral form as 
\begin{align}
\chi_{\mu\nu,\alpha\beta}^{\mathrm{R}}(\omega) &= \frac{1}{ZV}\sum_{mn} \frac{e^{-\beta E_n} - e^{-\beta E_m}}{\omega \hbar + i\delta + E_n - E_m}  \bra{n}J_{\mu,\alpha}\ket{m} \bra{m} J_{\nu,\beta} \ket{n}, \label{eq:retard-real-freq-spectral}\\
\chi_{\mu\nu,\alpha\beta}^{\mathrm{R}}(t) &= \frac{1}{ZV} \frac{-i}{\hbar} \theta(t)\sum_{nm} \left[e^{-\beta E_n} - e^{-\beta E_m}\right]\bra{n} J_{\mu,\alpha} \ket{m} \bra{m} J_{\nu,\beta} \ket{n} e^{i(E_n - E_m)t/\hbar}, \label{eq:retard-real-time-spectral}
\end{align}
where $Z$ is the partition function, and $E_n$ and $E_m$ denote the eigenvalues of the Hamiltonian.

By writing $\chi^{\mathrm{R}}_{\mu\nu,\alpha\beta}(\omega) =  \chi^{1}_{\mu\nu,\alpha\beta}(\omega) + i\chi^{2}_{\mu\nu,\alpha\beta}(\omega)$,
then using the Sokhotski-Plemelj theorem 
\begin{equation}
\frac{1}{x-x_0 + i0^+} = \mathrm{p.v.}\left(\frac{1}{x-x_0}\right) - i \pi \delta(x-x_0)
\end{equation}
we can break up \cref{eq:retard-real-freq-spectral} into
\begin{align}
\chi_{\mu\nu, \alpha \beta}^1(\omega) &= \frac{1}{ZV}\sum_{n m} \bra{n}J_{\mu,\alpha} \ket{m} \bra{m} J_{\nu,\beta} \ket{n} (e^{-\beta E_n} -  e^{-\beta E_m})\,\mathrm{p.v.}\left(\frac{1}{\omega + E_n - E_m}\right), \label{eq:chi1}\\
\chi_{\mu\nu,\alpha \beta}^2(\omega) &= \frac{-\pi}{ZV}\sum_{n m} \bra{n}J_{\mu,\alpha} \ket{m} \bra{m} J_{\nu,\beta} \ket{n} (e^{-\beta E_n} -  e^{-\beta E_m})\delta(\omega\hbar + E_n - E_m) \label{eq:chi2}.
\end{align}

DQMC measures unequal imaginary time (heat) current - (heat) current correlators
\begin{equation}
\chi_{\mu\nu,\alpha\beta}(\tau) =+\frac{1}{V}\langle J_{\mu,\alpha}(\tau) J_{\nu,\beta}(0)\rangle = \frac{+1}{ZV}\sum_{mn} e^{-\beta E_n} e^{\tau(E_n - E_m)}\bra{n} J_{\mu,\alpha} \ket{m} \bra{m} J_{\nu,\beta} \ket{n}\label{eq:matsubara-imag-time-spectral}.
\end{equation}
Comparing \cref{eq:chi2} with \cref{eq:matsubara-imag-time-spectral}, we find
\begin{equation}
\chi_{\mu\nu,\alpha\beta}(\tau) = \int_{-\infty}^{\infty} d(\omega \hbar) \frac{ e^{-\tau \omega \hbar}}{1-e^{-\beta \omega \hbar}}\frac{-\chi_{\mu\nu,\alpha\beta}^2(\omega)}{ \pi}.  \label{eq:maxent-relation}
\end{equation}
\cref{eq:maxent-relation} is the key relation that directly relates DQMC measurements in imaginary time to retarded correlators in real frequency. 

When $\mu=\nu$ and $\alpha = \beta$, we can show that \cref{eq:chi1} and \cref{eq:chi2} are purely real, and thus correspond to real and imaginary parts of $\chi^R_{\mu\mu, \alpha \alpha}$, respectively. 
This means that [c.f. \cref{eq:kubo}]
\begin{equation}
    \Im{L^{(\mu\mu)}_{\alpha\alpha}(\omega)} = \frac{\chi_{\mu\mu,\alpha\alpha}^1 (\omega)}{\omega}, \qquad \Re{L^{(\mu\mu)}_{\alpha\alpha}(\omega)} = \frac{-\chi_{\mu\mu,\alpha\alpha}^2 (\omega)}{\omega}.
\end{equation}
so that written out explicitly, we have
\begin{equation}
\chi_{11,xx}(\tau) = \frac{1}{V}\langle J_{x}(\tau) J_{x}(0)\rangle 
=  \int_{-\infty}^{\infty} d(\omega \hbar) \frac{ \omega e^{-\tau \omega \hbar}}{1-e^{-\beta \omega \hbar}}\frac{-\chi_{11,xx}^2(\omega)}{ \pi \omega}
=  \int_{-\infty}^{\infty} d(\omega \hbar) \frac{ \omega e^{-\tau \omega \hbar}}{1-e^{-\beta \omega \hbar}}\frac{\Re{L^{(11)}_{xx}(\omega)} }{ \pi} \label{eq:maxent-L11-diag}
\end{equation}
and
\begin{equation}
\chi_{22,xx}(\tau) = \frac{1}{V}\langle J_{Q,x}(\tau) J_{Q,x}(0) \rangle=  \int_{-\infty}^{\infty} d(\omega \hbar) \frac{ \omega e^{-\tau \omega \hbar}}{1-e^{-\beta \omega \hbar}}\frac{-\chi_{22,xx}^2(\omega)}{ \pi \omega}
=  \int_{-\infty}^{\infty} d(\omega \hbar) \frac{ \omega e^{-\tau \omega \hbar}}{1-e^{-\beta \omega \hbar}}\frac{\Re{L_{xx}^{(22)}(\omega)} }{ \pi}. \label{eq:maxent-L22-diag}
\end{equation}

We apply MaxEnt~\cite{Jarrell1996} analytic continuation to invert \cref{eq:maxent-L11-diag,eq:maxent-L22-diag}, and find the diagonal conductivities $\Re{L_{xx}^{(11)}(\omega)}$ and $\Re{L_{xx}^{(22)}(\omega)}$.

Experimentally, electric conductivity is typically measured under the condition $\nabla T = 0$, so $\sigma = L^{(11)}$, 
$\Re{\sigma_{xx}(\omega)} = \Re{L_{xx}^{(11)}(\omega)}$, and we obtain the DC value reported in main text by taking $\sigma^{\mathrm{DC}}_{xx} \equiv \Re{\sigma_{xx}(\omega\rightarrow 0)}$.

Thermal conductivity is typically measured under the zero electrical current condition $\vec{j} = 0$, so
\begin{equation}
    \kappa \equiv \kappa_{\mathrm{zc}} = \frac
    {1}{T} \left(L^{(22)} - L^{(21)} \left(L^{(11)}\right)^{-1} L^{(12)}\right) \label{eq:kappa-vs-L},
\end{equation}
where the first term may be called the nominal thermal conductivity corresponding to measurements under the condition $\vec{E}=0$,
\begin{equation}
    \kappa^0 = \frac{L^{(22)}}{T} \label{eq:nomimal-thermal-cond}.
\end{equation}
So $\Re{\kappa^0_{xx}(\omega)} = \Re{L_{xx}^{(22)}(\omega)}/T$, and we obtain the DC value reported in main text by taking $\kappa^{0,\mathrm{DC}}_{xx} \equiv \Re{\kappa^0_{xx}(\omega\rightarrow 0)}$. We do this to avoid inverting $L$ matrices with small elements, which will exacerbate statistical noise. The effect of the correction term in \cref{eq:kappa-vs-L} is small~\cite{Wang2022}.

On the other hand, when $J_{\mu,\alpha}\neq J_{\nu,\beta}$, \cref{eq:chi1} and \cref{eq:chi2} are not necessarily purely real, so do not necessarily correspond to real and imaginary parts of $\chi^R_{\mu\mu, \alpha \alpha}$. In the case of $\alpha = x$, $\beta = y$, and $\mu=\nu$, we can show that $\chi_{\mu\mu,x y}(\tau)$, $\chi_{\mu\mu,xy}^1(\omega)$, and $\chi_{\mu\mu,xy}^2(\omega)$ are all purely imaginary. Using $\chi_{\mu\mu,xy}(\tau)$ as an explicit example, we have
\begin{align}
    \langle J_{\mu,x}(\tau) J_{\mu,y}(0)\rangle &= \frac{1}{Z}\sum_{mn} e^{-\beta E_n} e^{\tau(E_n - E_m)}\bra{n} J_{\mu, x} \ket{m} \bra{m} J_{\mu,y} \ket{n} \\
    &= \frac{1}{Z}\sum_{mn} e^{-\beta E_n} e^{\tau(E_n - E_m)} \overline{\bra{n} J_{\mu,y} \ket{m}}\,\overline{\bra{m} J_{\mu, x} \ket{n}} =  \overline{\langle J_{\mu,y}(\tau) J_{\mu,x}(0)\rangle} = -\langle J_{\mu,y}(\tau) J_{\mu,x}(0)\rangle,
\end{align}
where the last equality used the C$_4$ symmetry of the square lattice. This means that [c.f. \cref{eq:kubo}]
\begin{equation}
    \Re{L^{(\mu\mu)}_{xy}(\omega)} = \frac{i\chi_{\mu\mu,xy}^1 (\omega)}{\omega}, \qquad \Im{L^{(\mu\mu)}_{xy}(\omega)} = \frac{i\chi_{\mu\mu,xy}^2 (\omega)}{\omega},
\end{equation}
so that written out explicitly, we have
\begin{equation}
-i\chi_{22,xy}(\tau) 
= \frac{-i}{V}\langle J_{Q,x}(\tau) J_{Q,y}(0)\rangle 
=  \int_{-\infty}^{\infty} d(\omega \hbar) \frac{ \omega e^{-\tau \omega \hbar}}{1-e^{-\beta \omega \hbar}}\frac{+i\chi_{22,xy}^2(\omega)}{ \pi \omega}
=  \int_{-\infty}^{\infty} d(\omega \hbar) \frac{ \omega e^{-\tau \omega \hbar}}{1-e^{-\beta \omega \hbar}}\frac{\Im{L^{(22)}_{xy}(\omega)} }{ \pi}\label{eq:maxent-L22-trans}.
\end{equation}

As a useful reference, \cref{tab:chi-symmetries} summarizes the properties of components of $\chi_{\mu\mu,\alpha\beta}^R(\omega) = \chi_{\mu\mu,\alpha\beta}^1(\omega) + i\chi_{\mu\mu,\alpha\beta}^2(\omega)$
\begin{table}[htb]
    \caption{Properties of components of $\chi_{\mu\mu,\alpha\beta}$}
    \label{tab:chi-symmetries}
    \centering

    \begin{tabular}{c|c|c}
    \textbf{component} & \textbf{real/imaginary?} & \textbf{symmetry} \\
    \hline \hline
    $\chi_{xx}^1(\omega)$ & real & even\\
    \hline
    $\chi_{xx}^2(\omega)$ & real & odd\\
    \hline
    $\chi_{xy}^1(\omega)$ & imaginary & odd\\
    \hline
    $\chi_{xy}^2(\omega)$ & imaginary & even\\

    \hline

    \hline
    \end{tabular}
\end{table}

The off-diagonal spectral weight $\chi_{\mu\mu,xy}^2(\omega)/\omega$ need not be positive over all frequencies, which precludes us from directly applying the standard MaxEnt algorithm to invert \cref{eq:maxent-L22-trans}. This is a known issue for off-diagonal spectral functions, and in this work, we use two strategies to tackle this, which we call the subtraction method (\cref{subsec:subtract}) and the proxy method (\cref{subsec:proxy}), respectively. 

\subsection{Subtraction method}
\label{subsec:subtract}
The subtraction method adopts the strategies of \cite{Reymbaut2015,Reymbaut2017}. Namely, we perform analytic continuation on the composite object $\chi_{\mu\mu,xx}(\tau) - i\chi_{\mu\mu,xy}(\tau)$, subtract out the longitudinal response $\chi_{\mu\mu,x x}^2(\omega)$ to obtain the transverse response $\chi_{\mu\mu,x y}^2(\omega)$, using the relation
\begin{align}
    \chi_{\mu\mu,xx}(\tau) - i\chi_{\mu\mu,xy}(\tau)  &= \frac{1}{V}\langle J_{\mu,x}(\tau) J_{\mu,x}(0)\rangle + \frac{-i}{V}\langle J_{\mu,x}(\tau) J_{\mu,y}(0)\rangle \nonumber \\
    &=\int_{-\infty}^{\infty} d(\omega \hbar) \frac{ \omega e^{-\tau \omega \hbar}}{1-e^{-\beta \omega \hbar}}\left[\frac{-\chi_{\mu\mu,xx}^2(\omega)}{ \pi \omega} + \frac{i\chi_{\mu\mu,xy}^2(\omega)}{ \pi \omega}\right] \label{eq:subtract-total}\\
    &=\int_{-\infty}^{\infty} d(\omega \hbar) \frac{ \omega e^{-\tau \omega \hbar}}{1-e^{-\beta \omega \hbar}}\left[\frac{\Re{L^{(\mu\mu)}_{xx}(\omega)} }{ \pi} + \frac{\Im{L^{(\mu\mu)}_{xy}(\omega)} }{ \pi}\right].
\end{align}
As long as the off-diagonal spectral weight is small, this procedure allows it to ``piggyback'' on a large positive diagonal spectral weight and allow MaxEnt to proceed as usual. This entails performing two MaxEnt fits, and subtracting them to obtain our desired result.

Because the analytic continuation relation \cref{eq:maxent-relation} only allows us to obtain $\chi^2_{\mu\mu,xy}(\omega)$ or $\Im{L^{(\mu\mu)}_{xy}(\omega)}$, we also need to perform a Kramers-Kronig transform after the subtraction of two MaxEnt spectra to obtain
\begin{equation}
    \Re{L^{(\mu\mu)}_{xy}(\omega)} = \frac{i\chi_{\mu\mu,xy}^1 (\omega)}{\omega} 
    = \frac{1}{\omega}\,\mathrm{p.v.}\int_{-\infty}^{\infty} 
    \frac{i\chi_{\mu\mu,xy}^2(\omega')}{\omega'-\omega}\frac{d\omega'}{\pi} 
    = \frac{1}{\omega}\,\mathrm{p.v.}\int_{-\infty}^{\infty} 
    \frac{\omega \Im{L^{(\mu\mu)}_{xy}(\omega)}}{\omega'-\omega}\frac{d\omega'}{\pi}.
\end{equation}
More specifically, we are interested in the DC value, obtained by
\begin{equation}
    \lim_{\omega \rightarrow 0}\Re{L^{(\mu\mu)}_{xy}(
    \omega)} =  i \,\mathrm{p.v.} \int \frac{d \omega'}{\pi} \frac{d}{d\omega}\frac{\chi_{\mu\mu,x y}^2(\omega')}{\omega' - \omega} \bigg|_{\omega = 0}=  i \int \frac{d\omega'}{\pi} \frac{\chi_{\mu\mu,x y}^2(\omega')}{\omega'^2}. \label{eq:kk-transform}
\end{equation}

\begin{figure}[htbp]
    \centering
    \includegraphics[width=\linewidth]{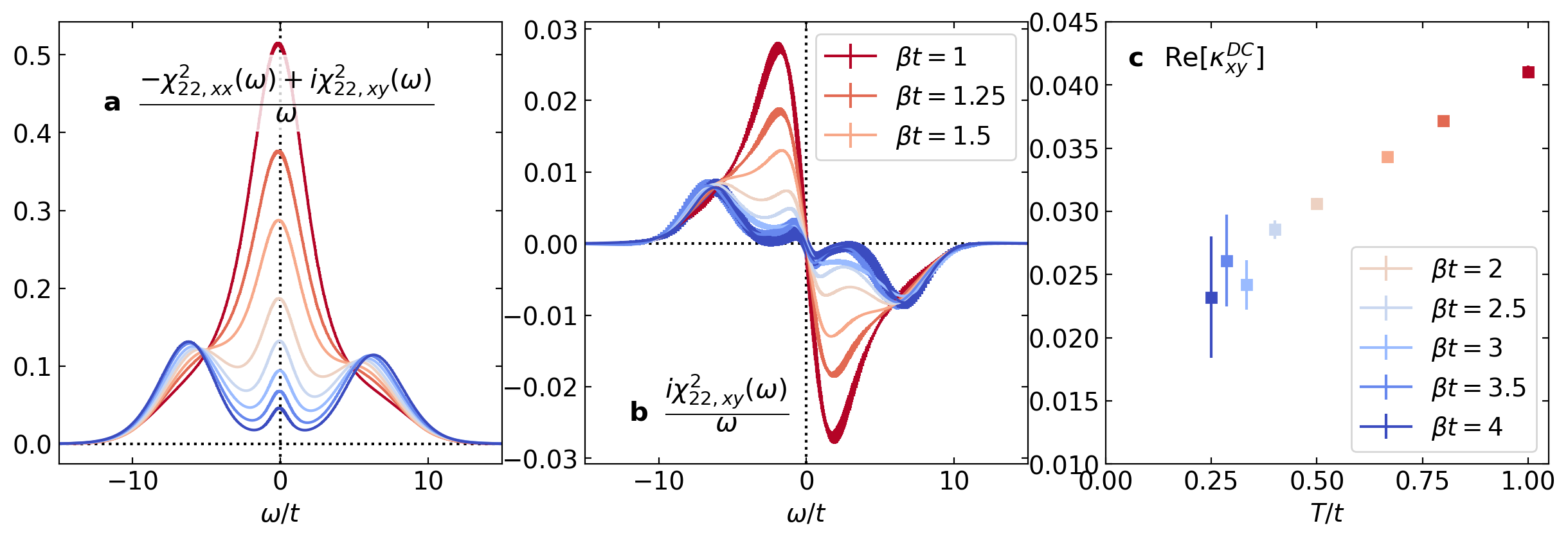}
    \caption[Examples of subtraction method for obtaining the thermal Hall conductivity]{Examples of MaxEnt subtraction method for obtaining the thermal Hall conductivity. \textbf{a} The result of analytically continuing the composite object, \cref{eq:subtract-total}, \textbf{b} The off-diagonal spectra, \textbf{c} the result after Kramers-Kronig transform, \cref{eq:kk-transform}. Next nearest neighbor hopping $t'/t = -0.1$, Hubbard $U/t = 6$, magnetic field strength $\Phi/\Phi_0 = 4/64$. Error bars denote one standard deviation of the mean, obtained via bootstrap resampling. }
    \label{fig:subtract-example}
\end{figure}

Some typical spectra and corresponding DC result obtained via this subtraction procedure are shown in \cref{fig:subtract-example}. 

Once we have obtained $\kappa_{xy}^{\mathrm{DC}} = \lim_{\omega \rightarrow 0}\Re{L^{(22)}_{xy}(\omega)}/T$ as outlined above, and $\kappa_{xx}^{\mathrm{DC}}$ via standard MaxEnt \cref{eq:maxent-L22-diag}, we can in turn calculate the thermal Hall angle $\theta_{\mathrm{th,H}}$, defined as 
\begin{equation}
\tan \theta_{\mathrm{th,H}} = \frac{\kappa^{\mathrm{DC}}_{xy}}{\kappa^{\mathrm{DC}}_{xx}}, \label{eq:def-angle-th-H}
\end{equation}
and the Hall coef{}ficient, $R_{\mathrm{th,H}} = \dfrac{\nabla T_y}{j_{q,x} B}$, defined as
\begin{equation}
R_{\mathrm{th,H}} = \frac{1}{B}\frac{\kappa^{\mathrm{DC}}_{xy}}{(\kappa^{\mathrm{DC}}_{xx})^2 + (\kappa^{\mathrm{DC}}_{xy})^2}. \label{eq:def-coeff-th-H}
\end{equation}

\subsection{Proxy method}
\label{subsec:proxy}

We have argued in an earlier work~\cite{Wang2021} about the properties of $\chi_{11,xy}(\tau)$; the situation is entirely analogous for thermal conductivity. $\chi_{22,xy}(\tau)$ is purely imaginary, and antisymmetric about $\tau = \beta/2$. 
By considering the Fourier transformed imaginary frequency correlator
\begin{equation}
\chi_{\mu\mu,\alpha\beta}(i\omega_n) = \int_0^\beta d\tau\, \chi_{\mu\mu,\alpha\beta}(\tau) e^{i\omega _n \tau}= \frac{1}{ZV}\sum_{mn} \frac{ e^{-\beta E_n} - e^{-\beta E_m}}{i\omega_n + E_n - E_m}\bra{n} J_{\mu,\alpha} \ket{m} \bra{m} J_{\nu,\beta} \ket{n}\label{eq:matsubara-imag-freq-spectral}
\end{equation}
we define \cite{Wang2021,Assaad1995}
\begin{align}
    L^{(\mu\mu)}_{xx}(i\omega_n) &= \frac{\chi_{\mu\mu,xx}(i\omega_n)-\chi_{\mu\mu,xx}(i\omega_n=0)}{\omega_n} \label{imaginaryfrequencyxx}, \\
    L^{(\mu\mu)}_{xy}(i\omega_n)&= \frac{\chi_{\mu\mu,xy}(i\omega_n)}{\omega_n}, \label{imaginaryfrequency} 
\end{align}
so that we obtain the finite-field version of Eq. (12) in \cite{Wang2021} for the thermal Hall coefficient: 
\begin{equation}
R_{\mathrm{th,H}}^{\mathrm{M1}}(i\omega_n) = \frac{1}{B} \frac{\chi_{22,xy}(i\omega_n) \omega_n T}{(\chi_{22,xx}(i\omega_n) - \chi_{22,xx}(0))^2 + \chi_{22,xy}(i\omega_n)^2}  \label{eq:proxy-thermal-m1}.
\end{equation}
This formula \cref{eq:proxy-thermal-m1} is exact for the DC thermal Hall coefficient at zero temperature,
\begin{equation}
    \lim_{i\omega_n \rightarrow 0} R_{\mathrm{th,H}}^{\mathrm{M1}}(i\omega_n) = \lim_{\omega\rightarrow 0} R_{\mathrm{th,H}}(\omega) = R_{\mathrm{th,H}}^{\mathrm{DC}}.
\end{equation}
At finite temperatures, we are only able to calculate \cref{eq:proxy-thermal-m1} for nonzero Matsubara frequencies, so we take the value of $R_{\mathrm{th,H}}^{\mathrm{M1}}(i\omega_n)$ at the lowest nonzero Matsubara frequency $\omega_1 = 2\pi/\beta$ as a proxy for $R_{\mathrm{th,H}}^{\mathrm{DC}}$. 
As long as $R_{\mathrm{th,H}}^{\mathrm{M1}}(i\omega_n)$ is well-behaved as a function of imaginary frequency, we have some confidence that its value at $\omega_0 = 0$ and $\omega_1 = 2\pi/\beta$ do not differ significantly.

\begin{figure}[htbp]
    \centering
    \includegraphics[width=\linewidth]{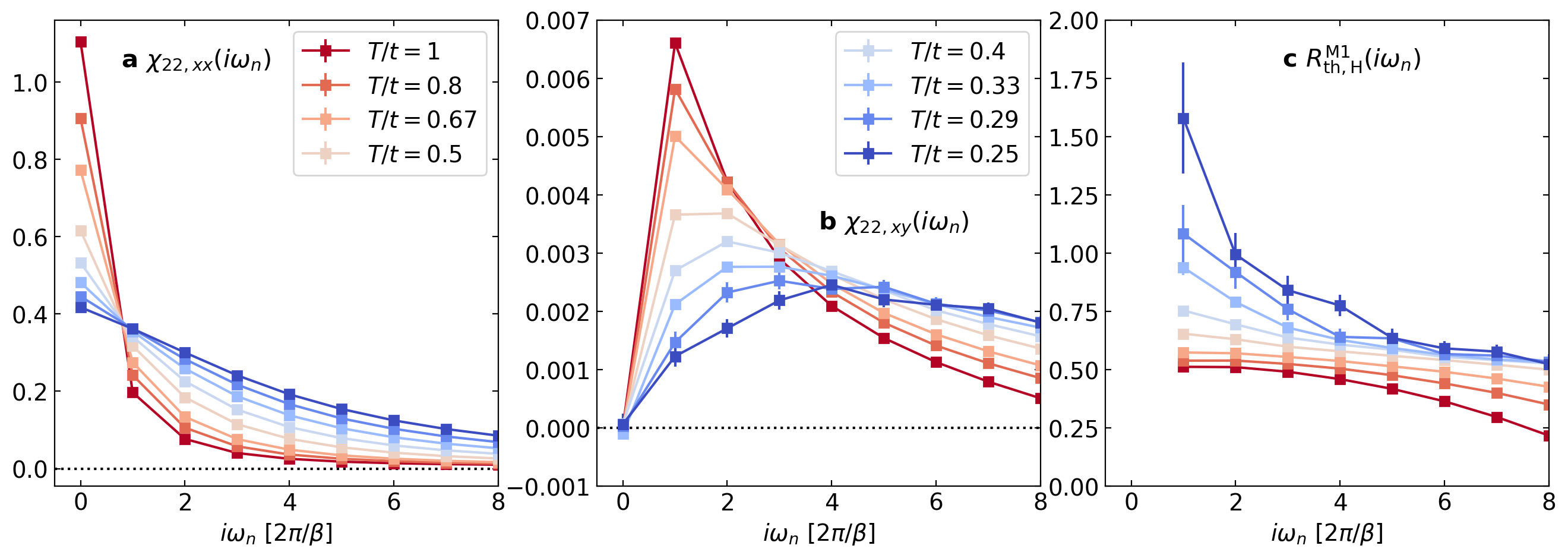}
    \caption[Examples of proxy method for obtaining the thermal Hall conductivity]{Examples of proxy method for obtaining the thermal Hall conductivity. Next nearest neighbor hopping $t'/t = -0.1$, Hubbard $U/t = 6$, magnetic field strength $\Phi/\Phi_0 = 1/64$. Error bars denote one standard deviation of the mean, obtained via jackknife resampling. }
    \label{fig:proxy-example}
\end{figure}

Some typical examples of $\chi_{22}$ and $R_{\mathrm{th,H}}$ results obtained via this proxy procedure are shown in \cref{fig:proxy-example}. 

Once we have obtained $R_{\mathrm{th,H}}$ as outlined above, and $\kappa_{xx}^{\mathrm{DC}}$ via standard MaxEnt \cref{eq:maxent-L22-diag}, we can use \cref{eq:def-angle-th-H,eq:def-coeff-th-H} to solve for  $\tan(\theta_{\mathrm{th,H}})$
\begin{equation}
    \frac{\tan(\theta_{\mathrm{th,H}})}{1+\tan^2(\theta_{\mathrm{th,H}})} = \kappa^{\mathrm{DC}}_{xx} \cdot B \cdot R_{\mathrm{th,H}}
\end{equation}
and derive
\begin{equation}
    \kappa^{\mathrm{DC}}_{xy} = \kappa^{\mathrm{DC}}_{xx} \tan(\theta_{\mathrm{th,H}})
\end{equation}

\section{Energy magnetization term}
\begin{figure}[htbp]
    \centering
    \includegraphics[width=0.5\columnwidth]{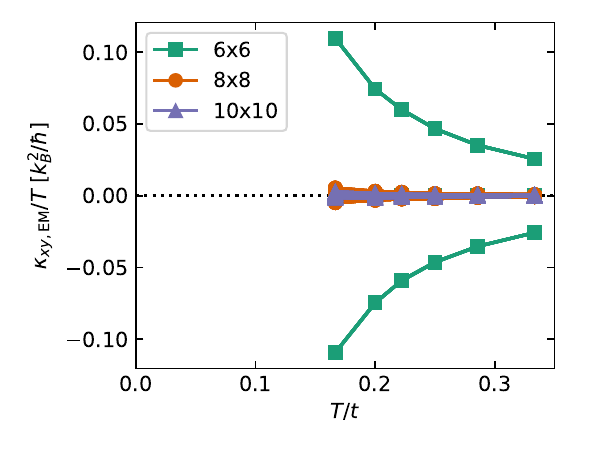}
    \caption{Finite size dependence of the energy magnetization correction $\kappa_{xy,\text{EM}}/T$, \cref{eq:energy-mag-correct}, calculated under periodic boundary conditions for $U/t=6$, $t'/t=-0.1$. The magnetic field strengths are $\Phi/\Phi_0=1/36$, $\Phi/\Phi_0=1/64$, and $\Phi/\Phi_0=1/100$ for $6\times 6$, $8\times 8$, and $10\times 10$ respectively. For each system size, $\kappa_{xy,\text{EM}}/T$ computed for all possible choices in origin are shown. }
    \label{fig:em_finite_size}
\end{figure}

\begin{figure}[htbp]
    \centering
    \includegraphics[width=0.7\columnwidth]{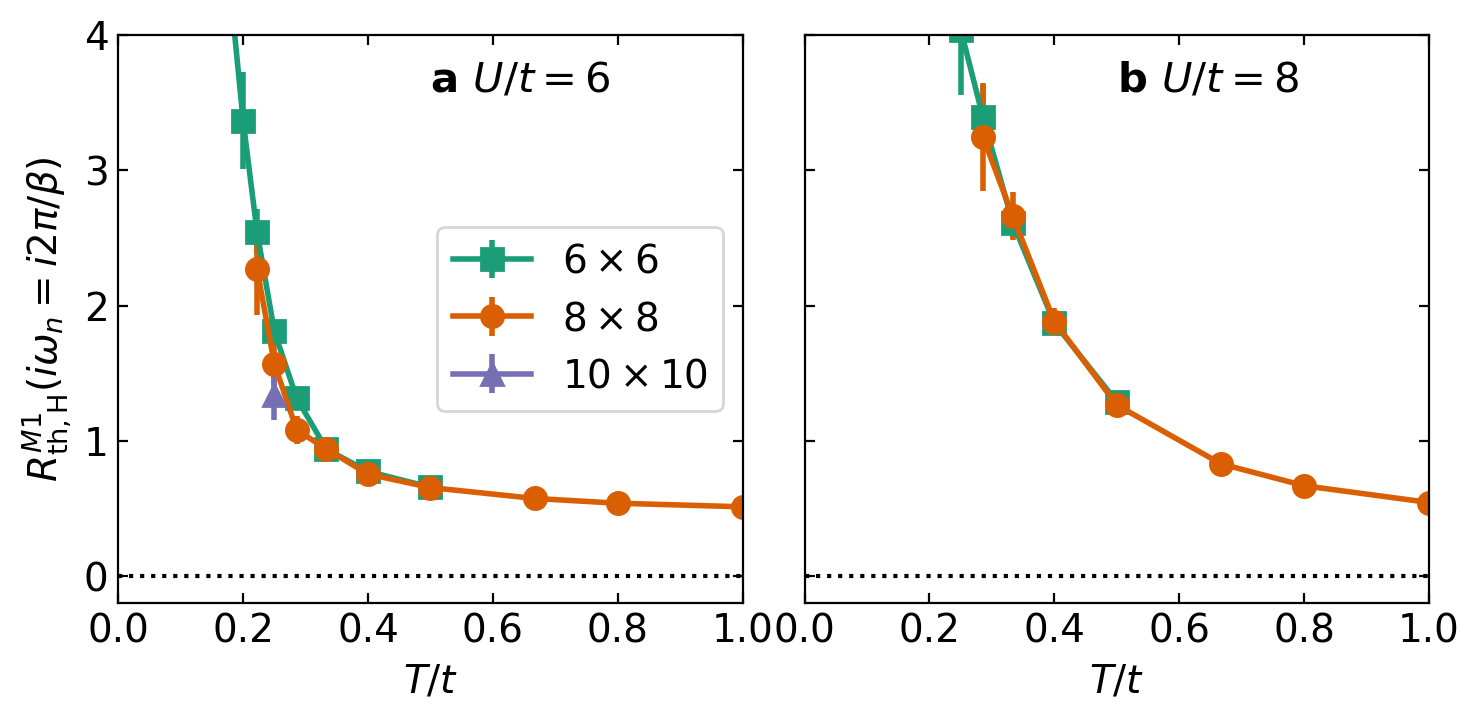}
    \caption{Finite size dependence of $R_{\mathrm{th,H}}^{\mathrm{M1}}$ proxy for Hubbard interaction \textbf{a} $U/t=6$ and \textbf{b} $U/t=8$, both with $t'/t = -0.1$ at half filling $\langle n \rangle = 1$. The magnetic field strength is $\Phi/\Phi_0 = 1/36$ on a $6\times 6$ cluster, $\Phi/\Phi_0 = 3/64$ on a $8\times 8$ cluster, and $\Phi/\Phi_0 = 2/100$ on a $10\times 10$ cluster}. Both panels share the same legend.
    \label{fig:proxy_finite_size}
\end{figure}

By introducing the pseudogravitational potential in \eqref{eq:pg}, there is an additional contribution to the heat current density $\mathbf{j}_{q}(\vec{r})$ stemming from the fact that $\psi$ couples to the energy density $h_i$ itself~\cite{Qin2011,Han2017}. In the DC limit, this extra contribution modifies the thermal conductivity by $\kappa_{xy}=\kappa_{xy,\text{Kubo}}+\kappa_{xy,\text{EM}}$, where $\kappa_{xy,\text{Kubo}}$ is the usual Kubo term discussed in \cref{eq:detailed-methods}, and $\kappa_{xy,\text{EM}}$ is the energy magnetization correction, given by 
\begin{align}
\kappa_{xy,\text{EM}} &= \frac{2}{TV} \int \text{d}\mathbf{r}\ \langle{r_{y} j_{Q,x}(\mathbf{r})\rangle}_0 = \frac{2}{T} \frac{1}{i}\Bigg\langle{ \left[\frac{\partial j_{Q,x}(\mathbf{q})}{\partial q_y} \right]_{\mathbf{q}=0}\Bigg\rangle}_0. \label{eq:energy-mag-correct}
\end{align}
Since we consider periodic boundary conditions, position $\vec{r}$ is not well-defined. In other words, the value of $\kappa_{xy,\text{EM}}$ in general depends on the choice in origin. Nonetheless, we see in Fig. \ref{fig:em_finite_size} that the overall magnitude of $\kappa_{xy,\text{EM}}/T$ systematically decreases as a function of system size regardless of choice in origin. In contrast, the proxy used to compute $\kappa_{xy,\text{Kubo}}/T$ does not drastically change as a function of system size, as shown in Fig. \ref{fig:proxy_finite_size}. Moreover, on a $8\times 8$ cluster, the maximum value for the energy magnetization correction is negligible compared to $\kappa_{xy,\text{Kubo}}/T$ at all temperatures. Thus, for the results in the main text, we only consider the Kubo contribution.

\section{Proof of zero transverse responses when \texorpdfstring{$t'=0$}{t'=0}}
\label{sec:proof-transverse}
Consider the Hubbard-Hofstadter Hamiltonian on a bipartite lattice at half filling, and define the charge conjugation transform as:
\begin{equation}
    c_{i\sigma} \rightarrow (-1)^i c_{i\sigma}^\dagger, \quad c_{i\sigma}^\dagger \rightarrow (-1)^i c_{i\sigma},\\
\end{equation}
where sign of prefactor $(-1)^i$ depending on sublattice. This is a unitary transformation. Without an applied magnetic field, the Hamiltonian is symmetric under charge conjugation: $H =  C H C^{-1}$. In the presence of a magnetic field, the Hamiltonian is not invariant under charge conjugation, but instead transforms under charge conjugation as
\begin{align}
    H(\vec{A}) \quad \rightarrow \quad  C H(\vec{A}) C^{-1} = H(-\vec{A}).
\end{align}
The electrical current operator \cref{eq:current-vector} and heat current operator \cref{heat-current} (with $\mu=0$ at half filling) transforms under charge conjugation as\footnote{We noticed an error in the supplementary material of Ref.~\cite{Wang2022}, where after the particle-hole transformation, the current operators should change as $J_K\to J_K$, $J_P\to J_P$, and $J\to -J$.}:
\begin{align}
    \vec{J}(\vec{A}) \quad \rightarrow \quad C\vec{J}(\vec{A})C^{-1} = -\vec{J}(-\vec{A})\\
    \vec{J}_Q(\vec{A}) \quad \rightarrow \quad C\vec{J}_Q(\vec{A})C^{-1} = \vec{J}_Q(-\vec{A}).
\end{align}
This means the transverse current-current correlator in imaginary time satisfies:
\begin{align}
    \chi_{11,xy}(\vec{A}) &= \expval{J_x(\vec{A},\tau) J_y(\vec{A})} = \text{Tr}\left[e^{-\beta H(\vec{A})}e^{\tau H(\vec{A})}  J_x(\vec{A}) e^{-\tau H(\vec{A})}  J_y(\vec{A})\right] \Big/ \text{Tr}\left[e^{-\beta H(\vec{A})}\right]\nonumber\\
    &= \text{Tr}\left[C e^{-\beta H(\vec{A})} e^{\tau H(\vec{A})} C^{-1} C J_x(\vec{A})  C^{-1} C e^{-\tau H(\vec{A})} C^{-1} C J_y(\vec{A}) C^{-1} \right] \Big/ \text{Tr}\left[C e^{-\beta H(\vec{A})} C^{-1}\right]\nonumber\\
    & = \text{Tr}\left[ e^{-\beta H(-\vec{A})} e^{\tau H(-\vec{A})}  J_x(-\vec{A})  e^{-\tau H(-\vec{A})} J_y(-\vec{A}) \right] \Big/ \text{Tr}\left[e^{-\beta H(-\vec{A})} \right]\nonumber\\ 
    &= \chi_{11,xy}(-\vec{A}). \label{eq:L11-xy-sym}
\end{align}
Analogously, the transverse heat current-heat current correlator in imaginary time obeys
\begin{align}
    \chi_{22,xy}(\vec{A}) = \chi_{22,xy}(-\vec{A}). \label{eq:L22-xy-sym}
\end{align}
For the sake of completeness, the charge current-heat current correlator in imaginary time obeys
\begin{align}
    \chi_{12,xx}(\vec{A}) &= \expval{J_x(\vec{A},\tau) J_{Q,x}(\vec{A})} = \text{Tr}\left[e^{-\beta H(\vec{A})}e^{\tau H(\vec{A})}  J_x(\vec{A}) e^{-\tau H(\vec{A})}  J_{Q,x}(\vec{A})\right] \Big/ \text{Tr}\left[e^{-\beta H(\vec{A})}\right] \nonumber \\
    &= \text{Tr}\left[C e^{-\beta H(\vec{A})} e^{\tau H(\vec{A})} C^{-1} C J_x(\vec{A})  C^{-1} C e^{-\tau H(\vec{A})} C^{-1} C J_{Q,x}(\vec{A}) C^{-1} \right] \Big/ \text{Tr}\left[C e^{-\beta H(\vec{A})} C^{-1}\right] \nonumber \\
    & = -\text{Tr}\left[ e^{-\beta H(-\vec{A})} e^{\tau H(-\vec{A})}  J_x(-\vec{A})  e^{-\tau H(-\vec{A})} J_{Q,x}(-\vec{A}) \right] \Big/ \text{Tr}\left[e^{-\beta H(-\vec{A})} \right] \nonumber \\ 
    &= -\chi_{12,xx}(-\vec{A}). \label{eq:L12-xx-sym}
\end{align}

In the presence of external magnetic field, The thermoelectric response coefficients obey the Onsager relations~\cite{Callen1948}
\begin{equation}
    L^{(\mu\nu)}_{\alpha\beta}(B) = L^{(\nu\mu)}_{\beta\alpha}(-B). \label{eq:onsager-relations}
\end{equation}
Combining \cref{eq:onsager-relations} with \cref{eq:L11-xy-sym,eq:L22-xy-sym,eq:L12-xx-sym} and Kubo formulas, we find that at half-filling, with $t' \neq 0 $, the Hubbard-Hofstadter Hamiltonian satisfies
\begin{equation}
    \alpha_{xx}(B) = \sigma_{xy}(B) = \kappa_{xy}(B) = 0.
\end{equation}

\section{Magnon-magnon scattering}

\subsection{Holstein-Primakoff expansion}
To study the effects of magnon-magnon interactions, we consider the low-energy effective spin model
\begin{equation}
     H_{\mathrm{eff}} = J_1 \sum_{\langle ij\rangle} \vec{S}_i  \cdot \vec{S}_j + J_2 \sum_{\langle \langle ij \rangle \rangle} \vec{S}_i  \cdot \vec{S}_j 
    + J_\chi \sum_{\triangle_{ijk}} \vec{S}_i \cdot (\vec{S}_j\times \vec{S}_k), \label{eq:spin-J1-J2-Jchi}
 \end{equation} 
 and perform a large $S$ expansion around a Neel state. We want to study perturbations of the spins away from the local $z$ axes, so we must first perform local rotations to each site. Taking $\mathbf{S}_{i}=R_{i}\tilde{\mathbf{S}}_{i}$ where $R\in $ SO(3), we can write the Hamiltonian in the rotated basis
\begin{equation}
H	=\sum_{i,j\in i}\tilde{\mathbf{S}}_{i}^{T}\tilde{H}_{ij}^{(2)}\tilde{\mathbf{S}}_{j}+\sum_{\triangle_{ijk}}\epsilon_{\alpha\beta\gamma}\tilde{S}_{i}^{\alpha}\tilde{S}_{j}^{\beta}\tilde{S}_{k}^{\gamma}\tilde{H}_{ijk}^{(3),\alpha\beta\gamma},\qquad\text{where }\tilde{H}_{ijk}^{(3),\alpha\beta\gamma}=R_{i}^{\alpha}R_{j}^{\beta}R_{k}^{\gamma}H_{ijk}^{(3),\alpha\beta\gamma},
\end{equation}
and $H^{(2)}_{ij}$ and $H^{(3)}_{ij}$ are tensors encoding the quadratic and cubic spin interactions. In this basis, we can perform local Holstein-Primakoff (HP) transformations
\begin{align}
\tilde{S}_{i}^{z}	&=S-a_{i}^{\dagger}a_{i}=S-n_{i} \\
\tilde{S}_{i}^{x}	&=\frac{\sqrt{2S-n_{i}}a_{i}+a_{i}^{\dagger}\sqrt{2S-n_{i}}}{2}\approx\sqrt{\frac{S}{2}}\left(a_{i}+a_{i}^{\dagger}\right) \\
\tilde{S}_{i}^{y}	&=\frac{\sqrt{2S-n_{i}}a_{i}-a_{i}^{\dagger}\sqrt{2S-n_{i}}}{2i}\approx-i\sqrt{\frac{S}{2}}\left(a_{i}-a_{i}^{\dagger}\right).
\end{align}
After performing the substitutions, the quadratic Hamiltonian is given by 
\begin{equation}
H^{(2)}=\sum_{\mathbf{k}}\varepsilon_{\mathbf{k}}\left(\alpha_{\mathbf{k}}^{\dagger}\alpha_{\mathbf{k}}+\beta_{\mathbf{k}}^{\dagger}\beta_{\mathbf{k}}\right),
\end{equation}
where $\alpha_{\mathbf{k}}$ and $\beta_{\mathbf{k}}$ are operators for the Bogoliubov quasiparticles, related to the magnon quasiparticles by 
\begin{align}
    a_{A,\mathbf{k}}	&=u_{\mathbf{k}}\alpha_{\mathbf{k}}+v_{\mathbf{k}}\beta_{-\mathbf{k}}^{\dagger}\\
    a_{B,\mathbf{k}}	&=u_{\mathbf{k}}\beta_{\mathbf{k}}+v_{\mathbf{k}}a_{-\mathbf{k}}^{\dagger}\\
    a_{A,\mathbf{k}}^{\dagger}	&=u_{\mathbf{k}}\alpha_{\mathbf{k}}^{\dagger}+v_{\mathbf{k}}\beta_{-\mathbf{k}}\\
    a_{B,\mathbf{k}}^{\dagger}	&=u_{\mathbf{k}}\beta_{\mathbf{k}}^{\dagger}+v_{\mathbf{k}}a_{-\mathbf{k}},
\end{align}
where $u_{\mathbf{k}}=\cosh\theta_{\mathbf{k}}$,   $v_{\mathbf{k}}=\sinh\theta_{\mathbf{k}}$, $\tanh2\theta_{\mathbf{k}}=-\gamma_{\mathbf{k}}$, and $\gamma_{\mathbf{k}}=\frac{1}{2}\left(\cos\left(k_{x}\right)+\cos\left(k_{y}\right)\right)$. The (degenerate) energies are given by 
\begin{equation}
\varepsilon_{\mathbf{k}}=4J_{1}S\phi_{\mathbf{k}}\sqrt{1-\left(\frac{\gamma_{\mathbf{k}}}{\phi_{\mathbf{k}}}\right)^{2}},
\end{equation}
where 
\begin{equation}
\phi_{\mathbf{k}}=1+\frac{J_{2}}{J_{1}}\left[\frac{1}{2}\left(\cos(k_{x}+k_{y})+\cos(k_{x}-k_{y})\right)-1\right].    
\end{equation}
Note that $J_{\chi}$ does not contribute to the quadratic Hamiltonian; rather, it  contributes quartic terms (to order $S$) to the Hamiltonian in the form of 
\begin{equation}
H^{(4)}=\sum_{\mathbf{k},\mathbf{k}_{1},\mathbf{k}_{2},\mathbf{k}_{3}}\delta_{\mathbf{k}+\mathbf{k}_{1}+\mathbf{k}_{2}+\mathbf{k}_{3}}W_{\mathbf{k},\mathbf{k}_{1},\mathbf{k}_{2},\mathbf{k}_{3}}^{\alpha\beta\gamma\delta}\psi_{\mathbf{k}}^{\alpha}\psi_{\mathbf{k}_{1}}^{\beta}\psi_{\mathbf{k}_{2}}^{\gamma}\psi_{\mathbf{k}_{3}}^{\delta},
\end{equation}
where $\psi_{\mathbf{k}}=\left(\alpha_{\mathbf{k}},\beta_{\mathbf{k}},\alpha_{\mathbf{-k}}^{\dagger},\beta_{\mathbf{-k}}^{\dagger}\right)$. The interaction vertices $W_{\mathbf{k},\mathbf{k}_{1},\mathbf{k}_{2},\mathbf{k}_{3}}^{\alpha\beta\gamma\delta}$ are terms of the form
\begin{equation}
\frac{iSJ_{\chi}}{N}\eta_{\mathbf{k}}\sum_{\left\{ \delta_{j},\delta_{k}\right\} \in\triangle}\left[\cos\left(\mathbf{q}_{1}\cdot\delta_{j}+\mathbf{q}_{2}\cdot\delta_{k}\right)\pm\cos\left(\mathbf{q}_{3}\cdot\delta_{j}+\mathbf{q}_{4}\cdot\delta_{k}\right)\right]    ,
\end{equation}
where $\mathbf{q}_{n}$ are linear combinations of $\mathbf{k},\mathbf{k}_{1},\mathbf{k}_{2},\mathbf{k}_{3}$,  $\delta_j$ and $\delta_k$ are displacement vectors to the $j$ and $k$ sites within a triangular plaquette, and $\eta_{\mathbf{k}}$ are various combinations of four $u_{\mathbf{k}}$ and $v_{\mathbf{k}}$  Since $u_{\mathbf{k}}$ and $v_{\mathbf{k}}$ are real, the vertices $W_{\mathbf{k},\mathbf{k}_{1},\mathbf{k}_{2},\mathbf{k}_{3}}^{\alpha\beta\gamma\delta}$ are purely imaginary. We may also consider quartic interactions at order $1/S$ coming from the $J_1$ and $J_2$ interactions of similar form, which are purely real. Many of these processes will contribute to longitudinal transport, but only some will contribute to the Hall transport. 

\subsection{Mean-field theory analysis}

First, we explore if spin interactions incorporated at the mean-field level to the magnon Hamiltonian is sufficient for ``escaping'' the no-go theorem and observing a finite thermal Hall signal.

We start with spin-wave theory on top of Neel order. The cubic spin interaction, which breaks TRS at the Hamiltonian level, is given as
\begin{align}
    H_{\chi} &=
  \frac{1}{2iS^3} \sum_{\triangle_{ijk}}  S_i^z (S_j^- S_k^+ - S_j^+ S_k^-) + S_j^z (S_k^- S_i^+ - S_k^+ S_i^-)+S_k^z (S_i^- S_j^+ - S_j^+ S_i^-) \\
  &=\frac{1}{iS^2}\sum_{j\in A} \sum_{\sigma,\sigma'\in \{\pm\}} \sigma \sigma' \left(\left[  a_{B,j-\sigma \hat x}^\dagger a_{B,j-\sigma \hat x} a_{A,j}^\dagger a_{B,j-\sigma' \hat y}^\dagger - H.c.\right] - (\hat x\leftrightarrow \hat y)\right) - \left(a_{A,j}^\dagger a_{A,j} a_{B,j-\sigma \hat y} a_{B,j-\sigma' \hat x}^\dagger - H.c.\right)\\
  &+\frac{1}{iS^2} \sum_{j\in B} (A\leftrightarrow B) ,
\end{align}
where we sum over all triangles with $i$, $j$, and $k$ labeling the vertices in a clockwise order. In the second line we have performed the HP transformation and dropped terms with more than four magnons, and $A$ and $B$ denote the $A$ and $B$ sublattice. There is no quadratic contribution from this term to the SWT Hamiltonian, as stated by the no-go theorem \cite{Katsura2010}.

We now see if the term contributes at the mean-field level. We perform the mean-field decoupling allowing as much freedom as possible. However, because the DQMC results see the thermal Hall effect even for extremely small $t'$, we assume that any signal is a feature of the Neel-ordered phase and therefore we should not break any symmetry present in the phase that is not explicitly broken by the perturbation. 

We define the mean-fields
\begin{equation}
    \Theta^{X_{\pm}Y_\pm}_{r'-r} = \langle a^{\pm}_{A,r} a^{\pm}_{B,r'}\rangle
\end{equation}
e.g. $\Theta^{A_+ B_-}_x = \langle a^\dagger_{A,j} a_{B,j+x}\rangle$. Due to not wanting to break additional symmetry, we only allow for magnetization conserving terms: $a_{A,i}^\dagger a_{A,j}$, $a_{B,i}^\dagger a_{B,j}$, $a_{A,i} a_{B,j}$, and $a_{A,i}^\dagger a_{B,j}^\dagger$. With this restriction, the only mean-fields that appear are
\begin{equation}\begin{aligned}
    \Theta^{A_+,B_+}_{nn} &= \Theta_{\hat x}^{A_+ B_+}=\Theta_{-\hat x}^{A_+ B_+}=\Theta_{\hat y}^{A_+ B_+}=\Theta_{-\hat y}^{A_+ B_+}; \\
    \Theta^{B_-B_+}_{nnn} &= \Theta^{B_-B_+}_{\hat y+\hat x}=\Theta^{B_-B_+}_{\hat y-\hat x}; \qquad \Theta^{A_-A_+}_{nnn} = \Theta^{A_-A_+}_{\hat y+\hat x}=\Theta^{A_-A_+}_{\hat y-\hat x}; \qquad
    \Delta S = \Theta^{A_+ A_-}_0, \Theta^{B_+ B_-}_0;
\end{aligned}
\end{equation}
where we made use of $C_4$ symmetry to group terms that all should take the same value. Note that $(\Theta_{nnn}^{X_-X_+})^*=(\Theta_{\hat y + \hat x}^{B_-B_+})^* = \Theta^{B_-B_+}_{-\hat y-\hat x}= \Theta_{\hat y + \hat x}^{B_-B_+}$ using $C_4$ symmetry. 

After mean-field decoupling the Hamiltonian and performing a Fourier transform, we arrive at
\begin{equation}
\begin{aligned}
    iS^2 H_\chi &= C_{\chi} + \sum_{\pmb k} Q_{AA}(\pmb k) a_{A,\pmb k}^\dagger a_{A,\pmb k} +Q_{BB}(\pmb k) a_{B,\pmb k}^\dagger a_{B,\pmb k} + Q_{AB}(\pmb k) a_{A,-\pmb k} a_{B,\pmb k} + Q_{AB}(\pmb k)^* a_{A,-\pmb k}^\dagger a_{B,\pmb k}^\dagger \\
       Q_{AA}(\pmb k) &=\sum_{\sigma \sigma'} \sigma \sigma'\left[\Theta_{-\sigma \hat x}^{A_- B_-} e^{i\pmb k\cdot (\sigma' \hat y - \sigma \hat x)} - \Theta_{-\sigma \hat x}^{A_+ B_+} e^{i\pmb k \cdot (\sigma \hat x- \sigma' \hat y)} + \Theta_{-\sigma' \hat y}^{A_+ B_+} e^{i\pmb k \cdot (\sigma' \hat y - \sigma \hat x)} - \Theta_{-\sigma' \hat y}^{A_- B_-} e^{i\pmb k \cdot (\sigma \hat x - \sigma' y)} \right]=0\\
   Q_{BB}(\pmb k) &=\sum_{\sigma \sigma'}\sigma\sigma'\left[ \Theta_{-\sigma \hat x}^{B_- A_-} e^{i\pmb k \cdot (\sigma' \hat y - \sigma \hat x)} - \Theta_{-\sigma \hat x}^{B_+ A_+} e^{i\pmb k \cdot (\sigma \hat x- \sigma' \hat y)} + \Theta_{-\sigma'\hat y}^{B_+ A_+} e^{i\pmb k \cdot (\sigma' \hat y - \sigma \hat x)} - \Theta_{-\sigma' \hat y}^{B_- A_-} e^{i\pmb k \cdot (\sigma \hat x - \sigma' \hat y)}\right]=0 \\
   Q_{AB}(\pmb k) &= \sum_{\sigma \sigma'} \sigma \sigma' \Big[e^{ik_x\sigma} (\Theta_{\sigma' \hat y-\sigma x}^{B_+ B_-} - \Theta_{\sigma' \hat y}^{A_+ B_+}) + e^{ik_y\sigma'} (\Theta_{\sigma \hat x}^{A_+ B_+} - \Theta_{\sigma \hat x - \sigma' \hat y}^{B_+ B_-}) \\
   &+ e^{-ik_x\sigma } ( \Theta_{\sigma'\hat y - \sigma \hat x}^{A_+ A_-} - \Theta_{\sigma' \hat y}^{B_+ A_+}) + e^{-ik_y \sigma'}(\Theta_{\sigma \hat x}^{B_+ A_+} - \Theta_{\sigma \hat x - \sigma' \hat y}^{A_+ A_-})\Big] \\
   &=\sum_{\sigma,\sigma'}\sigma \sigma' \Big[e^{ik_x \sigma } (\Theta_{nnn}^{B_+ B_-} - \Theta_{nnn}^{A_+ A_-}) + e^{ik_y\sigma'} (\Theta_{nnn}^{A_+ A_-} - \Theta_{nnn}^{B_+ B_-})  \Big]=0 
\end{aligned}
\end{equation}
where $C_\chi=0$ through a similar manipulations. We used that $\Theta_{nnn}^{B_+ B_-} = \Theta_{nnn}^{A_+A_-}$, since, if there were spontaneous sublattice symmetry breaking, it would be detectable in DQMC in long-range correlators, but the numerical results preserve translation symmetry. Therefore there is no contribution to the quadratic magnon Hamiltonian from $H_\chi$ at the mean-field level. This derivation excludes an emergent Berry curvature explanation for the observed thermal Hall signal and instead suggests that magnon scattering as the only possible mechanism.

\subsection{Time Reversal Symmetry}
Before we proceed with the scattering theory, we first examine the time-reversal symmetry (TRS) of the Bogoliubov Hamiltonian since breaking TRS allows for a finite thermal Hall coefficient. We emphasize here that TRS is defined with respect to the Bogoliubov operators rather than the usual spin operators. Although a spin Hamiltonian may (may not) be TRS, the corresponding magnon Hamiltonian can break (preserve) TRS. For example, a Zeeman field breaks TRS in the spin language, but is preserved in HP operators. The TR operator $\mathcal{T}$ in the HP language can be defined as
\begin{align}
\mathcal{T}\alpha_{\mathbf{k}}\mathcal{T}^{-1}&=\alpha_{-\mathbf{k}}\\
\mathcal{T}\alpha_{\mathbf{k}}^{\dagger}\mathcal{T}^{-1}&=\alpha_{-\mathbf{k}}^{\dagger}\\
\mathcal{T}\beta_{\mathbf{k}}\mathcal{T}^{-1}&=\beta_{-\mathbf{k}}\\
\mathcal{T}\beta_{\mathbf{k}}^{\dagger}\mathcal{T}^{-1}&=\beta_{-\mathbf{k}}^{\dagger}.
\end{align}
Since $\mathcal{T}$ is antiunitary, it also contains complex conjugation $\mathcal{T} c \mathcal{T}^{-1}=c^*$. The Hamiltonian is TR invariant if $H=\mathcal{T} H\mathcal{T}^{-1}$.
Thus, the conditions for our Hamiltonian to be TR symmetric are:
\begin{align*}
H^{(2)} & :\qquad\varepsilon_{\mathbf{k}}=\varepsilon_{-\mathbf{k}}\\
H^{(4)} & :\qquad\left(W_{\mathbf{k},\mathbf{k}_{1},\mathbf{k}_{2},\mathbf{k}_{3}}^{\alpha\beta\gamma\delta}\right)^{*}=W_{\mathbf{-k},\mathbf{-k}_{1},\mathbf{-k}_{2},\mathbf{-k}_{3}}^{\alpha\beta\gamma\delta}.
\end{align*}
In our case, $\left(W_{\mathbf{k},\mathbf{k}_{1},\mathbf{k}_{2},\mathbf{k}_{3}}^{\alpha\beta\gamma\delta}\right)^{*}=-W_{\mathbf{-k},\mathbf{-k}_{1},\mathbf{-k}_{2},\mathbf{-k}_{3}}^{\alpha\beta\gamma\delta}$, for the vertices involving the $J_\chi$, thus it is possible to generate a finite thermal Hall
effect. 
\subsection{Semi-Classical Boltzmann Transport}
We use semi-classical Boltzmann transport theory to compute the thermal
Hall due to magnon-magnon scattering~\cite{Chatzichrysafis2024}.
From Fourier's law (i.e., a restatement of Eq. \ref{eq:onsager-2}), the thermal conductivity is related to the magnon heat current by 
\begin{equation}
    j_{q}^{\mu}=-\kappa^{\mu\nu}\partial_{\nu}T.
\end{equation}
The magnon heat current density is given by 
\begin{align}
\mathbf{j}_{q} & =\frac{1}{V}\sum_{\mathbf{k}}\varepsilon_{\mathbf{k}}\mathbf{v}_{\mathbf{k}}N_{\mathbf{k}},
\end{align}
where $\varepsilon_{\mathbf{k}}$ is the free magnon dispersion, $\mathbf{v}_{\mathbf{k}}=\partial\varepsilon_{\mathbf{k}}/\partial\mathbf{k}$
is the magnon group velocity, and $V$ is the volume of the system.
$N_{\mathbf{k}}=N_{\mathbf{k}}\left(t,\mathbf{r}(t)\right)$ is the
out of equilibrium magnon distribution function and can be computed using the semi-classical Boltzmann equation (BE) 
\begin{equation}
    \mathbf{v}_{\mathbf{k}}\cdot\nabla T\frac{\partial N_{\mathbf{k}}}{\partial T}=I_{\mathbf{k}}^{\text{coll}},
\end{equation}
where $I_{\mathbf{k}}^{\text{coll}}$ is the collision integral. The left hand side represents the diffusion due to the temperature gradient and the right hand side represents the scattering rate between the magnons mediated by
magnon-magnon interactions. The collision integral is
\begin{equation}
    I_{\mathbf{k}}^{\text{coll}}=\sum_{\mathbf{k}'}\Gamma_{\mathbf{k}}\left[\left\{ N_{\mathbf{k}'}\right\} \right]=\sum_{\mathbf{k}'}\left(\Gamma_{\mathbf{k}}^{\text{in}}\left[\left\{ N_{\mathbf{k}'}\right\} \right]-\Gamma_{\mathbf{k}}^{\text{out}}\left[\left\{ N_{\mathbf{k}'}\right\} \right]\right),
\end{equation}
where $\Gamma_{\mathbf{k}}\left[\left\{ N_{\mathbf{k}'}\right\} \right]$
are all the scattering processes due to magnons with a momentum $\mathbf{k'}$.
We can compute $\Gamma_{\mathbf{k}}$ using Fermi's golden rule, or 
\begin{equation}
\Gamma_{\text{if}}\left[\left\{ N_{\mathbf{k}'}\right\} \right]=\frac{2\pi}{\hbar}\abs{T_{\text{if}}}^{2}\delta\left(E_{i}-E_{f}\right),
\end{equation}
where the transition matrix $T_{\text{if}}$ is given by 
\begin{align}
T_{\text{if}}&=\bra fT\ket i , \\ 
T&=H_{\text{int}}+H_{\text{int}}\left(\sum_{\nu}\frac{\ket{\nu}\bra{\nu}}{E_{i}-E_{\nu}+i\eta}\right)H_{\text{int}}+\dots,\qquad\eta>0
\end{align}
and $\nu$ labels the intermediate states. We make use of the relations
\begin{align*}
\alpha_{\mathbf{k}}^{\dagger}(\beta_{\mathbf{k}}^{\dagger})\ket{\dots,N_{\mathbf{k}},\dots} & =\sqrt{N_{\mathbf{k}}+1}\ket{\dots,N_{\mathbf{k}},\dots}\\
\alpha_{\mathbf{k}}(\beta_{\mathbf{k}})\ket{\dots,N_{\mathbf{k}},\dots} & =\sqrt{N_{\mathbf{k}}}\ket{\dots,N_{\mathbf{k}},\dots}.
\end{align*}
Since the energies are degenerate, we omit differentiating between the two types of magnon distribution functions. After obtaining the scattering rate, we linearize it with respect to $\delta N_{\mathbf{k}}$,
where $N_{\mathbf{k}}=\bar{N}_{\mathbf{k}}+\delta N_{\mathbf{k}}$, $\bar{N}_{\mathbf{k}}$ is the Bose-Einstein distribution, and
$\delta N_{\mathbf{k}}$ is the out-of-equilibrium distribution. This
amounts to writing 
\begin{equation}
\Gamma_{\mathbf{k}}\left[\left\{ N_{\mathbf{k}_{i}}\right\} \right]=\sum_{\mathbf{k}'}C_{\mathbf{kk'}}\delta N_{\mathbf{k}'},
\end{equation}
where $C_{\mathbf{kk'}}$ is a matrix containing the collision kernel.
The diagonal elements of $C_{\mathbf{kk'}}$ are denoted by $D_{\mathbf{k}}$,
which is also the inverse relaxation time $D_{\mathbf{k}}=\tau_{\mathbf{k}}^{-1}$.
One can also introduce a phenomenological damping term into $D_{\mathbf{k}}$,
but we do not consider such a term here. The off-diagonal elements
are denoted by $O_{\mathbf{kk'}}$, and are the elements that will
contribute to the thermal Hall conductivity. We switch to Hardy's basis, such that $\mathcal{C_{\mathbf{kk'}}}=\frac{G_{\mathbf{k'}}}{G_{\mathbf{k}}}C_{\mathbf{kk'}}$, 
where $G_{\mathbf{k}}=\sqrt{\bar{N}_{\mathbf{k}}\left(\bar{N}_{\mathbf{k}}+1\right)}$.
The thermal conductivity tensor is given by 
\begin{equation}
    \kappa_{\mathrm{th}}=-\frac{1}{k_{B}T^{2}V}\sum_{\mathbf{kk'}}\mathbf{v}_{\mathbf{k}}\otimes\mathbf{v}_{\mathbf{k'}}\varepsilon_{\mathbf{k}}\varepsilon_{\mathbf{k'}}G_{\mathbf{k}}G_{\mathbf{k}'}\left[\mathcal{C}^{-1}\right]_{\mathbf{kk'}}.
\end{equation}

After symmetrizing, the thermal Hall conductivity is given by 
\begin{equation}
    \kappa_{\mathrm{th,H}}=\frac{1}{2k_{B}T^{2}V}\sum_{\mathbf{k,k'}}(\mathbf{v}_{\mathbf{k}}\times\mathbf{v}_{\mathbf{k}'})_z\varepsilon_{\mathbf{k}}\varepsilon_{\mathbf{k}'}\tau_{\mathbf{k}}\tau_{\mathbf{k'}}G_{\mathbf{k}}G_{\mathbf{k'}}\left(\frac{\mathcal{O}_{\mathbf{kk'}}-\mathcal{O}_{\mathbf{k'k}}}{2}\right),
\end{equation}
where 
\begin{equation}
\mathcal{O}_{\mathbf{kk'}} =\mathcal{O}_{\mathbf{kk'}}^{\text{in}}-\mathcal{O}_{\mathbf{kk'}}^{\text{out}}=\left(\mathcal{O}_{\mathbf{kk'}}^{++}+\mathcal{O}_{\mathbf{kk'}}^{+-}\right)-\left(\mathcal{O}_{\mathbf{kk'}}^{-+}+\mathcal{O}_{\mathbf{kk'}}^{--}\right).
\end{equation}
Then the antisymmetric part of the collision
kernel is given by 
\begin{equation}
\left(\frac{\mathcal{O}_{\mathbf{kk'}}-\mathcal{O}_{\mathbf{k'k}}}{2}\right) =\frac{\bar{N}_{\mathbf{k}'}-\bar{N}_{\mathbf{k}}}{2\bar{N}_{\mathbf{k}'}}\left(\mathcal{O}_{\mathbf{kk'}}^{++}-e^{-\beta\varepsilon_{\mathbf{k}'}}\mathcal{O}_{\mathbf{kk'}}^{--}\right)+\frac{\bar{N}_{\mathbf{k}'}+\bar{N}_{\mathbf{k}}+1}{2\bar{N}_{\mathbf{k}'}}\left(e^{-\beta\varepsilon_{\mathbf{k}'}}\mathcal{O}_{\mathbf{kk'}}^{+-}-\mathcal{O}_{\mathbf{kk'}}^{-+}\right).
\end{equation}
The key point here is that the terms that will contribute something finite to $\kappa_{\mathrm{th,H}}$ are the terms that break the microscopic detailed-balance conditions, as mentioned in the main text. 

\subsection{Scattering from interactions}

We may depict the different scattering processes using Feynman diagrams. We choose the convention that an arrow going into a vertex creates a quasiparticle in the diagonal basis, and an arrow leaving a vertex annihilates one. The scattering transition probabilities $\abs{T_{\text{if}}}^{2}$ result in a change in net magnon number $\Delta N=\left\{ -4,-2,0,2,+4\right\} $. 

Due to the Bogoliubov transformation, we have many 
vertices, thus we group them in terms of net magnon number. Each scattering channel will contribute something
positive definite, thus we examine one process in the $\Delta N= 0 $ channel as an example. We consider a 2-in-2-out process  
\begin{equation}
    \ket{\text{i}}=\ket{N_{\mathbf{k}},N_{\mathbf{k}_{1}},N_{\mathbf{k}_{2}},N_{\mathbf{k}_{3}}}
    \qquad\ket{\text{f}}=\ket{N_{\mathbf{k}}-1,N_{\mathbf{k}_{1}}-1,N_{\mathbf{k}_{2}}+1,N_{\mathbf{k}_{3}}+1}
\end{equation}
so the transition probability is
\begin{equation}
\left|T_{\text{if}}\right|^{2}=\left|T_{\text{if}}^{(1)}+T_{\text{if}}^{(2)}\right|^{2}=
\left|
\vcenter{\hbox{
\begin{tikzpicture}[scale=0.5,transform shape]
\begin{feynman}
	\vertex (a){\(\mathbf{k}\)};
	\vertex [dot,below right=of a](b){};
	\vertex [below left=of b](c){\(\mathbf{k}_1\)};
	\vertex [above right=of b](d){\(\mathbf{k}_2\)};
	\vertex [below right=of b](e){\(\mathbf{k}_3\)};
	
	\diagram*{
		(a) -- [fermion] (b) -- [fermion] (d),
		(c) -- [fermion] (b) -- [fermion] (e), 
	};
\end{feynman}
\end{tikzpicture}
}}
+
\vcenter{\hbox{
\begin{tikzpicture}[scale=0.5,transform shape]
\begin{feynman}
	\vertex (a){\(\mathbf{k}\)};
	\vertex [dot,below right=of a](b){};
	\vertex [below left=of b](c){\(\mathbf{k}_1\)};
    \vertex [dot,right=of b](w){};
	\vertex [above right=of w](d){\(\mathbf{k}_2\)};
	\vertex [below right=of w](e){\(\mathbf{k}_3\)};
	\diagram*{
		(a) -- [fermion] (b) -- [anti fermion] (c), 
        (b) -- [fermion, half left] (w),
        (w) -- [fermion, half left] (b),
        (d) -- [anti fermion] (w) -- [fermion] (e),
	};
\end{feynman}
\end{tikzpicture}
}}
+\dots
\right|^2,
\end{equation}
where the $\dots$ correspond to other tree and bubble diagrams with $\Delta N=0$. 

From Fermi's golden rule, the scattering rates can be calculated with
\begin{align}
\Gamma_{\text{if}}^{(1)}\left[\left\{ N_{\mathbf{k}'}\right\} \right] & =\frac{2\pi}{\hbar}\abs{T_{\text{if}}^{(1)}}^{2}\delta\left(E_{\text{f}}-E_{\text{i}}\right)\\
\Gamma_{\text{if}}^{(2)}\left[\left\{ N_{\mathbf{k}'}\right\} \right] & =\frac{2\pi}{\hbar}\abs{T_{\text{if}}^{(2)}}^{2}\delta\left(E_{\text{f}}-E_{\text{i}}\right)\\
\Gamma_{\text{if}}^{\text{(1,2)}}\left[\left\{ N_{\mathbf{k}'}\right\} \right] & =\frac{4\pi}{\hbar}\delta\left(E_{\text{f}}-E_{\text{i}}\right)\left[\text{Re}\left(T_{\text{if}}^{(1)}\right)\text{Re}\left(T_{\text{if}}^{(2)}\right)+\text{Im}\left(T_{\text{if}}^{(1)}\right)\text{Im}\left(T_{\text{if}}^{(2)}\right)\right].
\end{align}
Note that both $\Gamma_{\text{if}}^{(1)}$ and $\Gamma_{\text{if}}^{(2)}$
are positive definite, whereas $\Gamma_{\text{if}}^{(1,2)}$ can be
either positive or negative. However, since $\Gamma_{\text{if}}^{(1)}\sim O(J_{\chi}^{2})$,
$\Gamma_{\text{if}}^{(1,2)}\sim O(J_{\chi}^{3})$, $\Gamma_{\text{if}}^{(2)}\sim O(J_{\chi}^{4})$, then 
$\Gamma_{\text{if}}^{(1)}>\Gamma_{\text{if}}^{\text{(1,2)}}>\Gamma_{\text{if}}^{(2)}$. Thus, although $\Gamma_{\text{if}}^{(1,2)}$ can be positive or negative,
the total scattering rate $\Gamma_{\text{if}}>0$ since it will be $\Gamma_{\text{if}}^{(1)}\pm$
a small number (otherwise the perturbation theory breaks down).

\subsubsection{First order contribution}
At first order in $J_\chi$ and $S$, the out-scattering process (2-in-2-out, denoted by +2 for 2 magnons out) is given by 
\begin{align}
\Gamma_{\mathbf{k}}^{(1),\text{out}} & =\frac{2\pi}{\hbar}
\sum_{\mathbf{k}_{1},\mathbf{k}_{2},\mathbf{k}_{3}}
\delta\left(\varepsilon_{\mathbf{k}}+\varepsilon_{\mathbf{k}_{1}} -\varepsilon_{\mathbf{k}_{2}}-\varepsilon_{\mathbf{k}_{3}}\right)
\delta_{\mathbf{k}+\mathbf{k}_{1}-\mathbf{k}_{2}-\mathbf{k}_{3}}
\abs{W_{\mathbf{k},\mathbf{k}_{1},\mathbf{k}_{2},\mathbf{k}_{3}}^{+2}}^{2}
N_{\mathbf{k}} N_{\mathbf{k}_{1}}\left(N_{\mathbf{k}_{2}}+1\right)\left(N_{\mathbf{k}_{3}}+1\right),
\end{align}
and the in-scattering process (2-out-2-in, denoted by -2 for two magnons in) is given by 
\begin{align}
\Gamma_{\mathbf{k}}^{(1),\text{in}} & =\frac{2\pi}{\hbar}
\sum_{\mathbf{k}_{1},\mathbf{k}_{2},\mathbf{k}_{3}}
\delta\left(\varepsilon_{\mathbf{k}}+\varepsilon_{\mathbf{k}_{1}}-\varepsilon_{\mathbf{k}_{2}}-\varepsilon_{\mathbf{k}_{3}}\right)
\delta_{\mathbf{k}+\mathbf{k}_{1}-\mathbf{k}_{2}-\mathbf{k}_{3}}
\abs{W_{\mathbf{k},\mathbf{k}_{1},\mathbf{k}_{2},\mathbf{k}_{3}}^{+2}}^{2}(N_{\mathbf{k}} +1) (N_{\mathbf{k}_{1}}+1)N_{\mathbf{k}_{2}}N_{\mathbf{k}_{3}},
\end{align}
where we used $W_{\mathbf{k},\mathbf{k}_{1},\mathbf{k}_{2},\mathbf{k}_{3}}^{-2}=\left(W_{\mathbf{k},\mathbf{k}_{1},\mathbf{k}_{2},\mathbf{k}_{3}}^{+2}\right)^{*}$.
Next, we linearize $\Gamma_{\mathbf{k}}^{(1)}=\Gamma_{\mathbf{k}}^{(1),\text{in}}-\Gamma_{\mathbf{k}}^{(1),\text{out}}$
such that 
\begin{equation}
    \Gamma_{\mathbf{k}}^{(1)}=C\left[\left\{ \bar{N_{\mathbf{k}}},\bar{N_{\mathbf{k}_{1}}},\bar{N_{\mathbf{k}_{2}}},\bar{N_{\mathbf{k}_{3}}}\right\} \right]+D_{\mathbf{k}}\delta N_{\mathbf{k}}+\sum_{\mathbf{k}_{1}}O_{\mathbf{k},\mathbf{k}_{1}}^{\text{}}\delta N_{\mathbf{k}_{1}}+\sum_{\mathbf{k}_{2}}O_{\mathbf{k},\mathbf{k}_{2}}^{\text{}}\delta N_{\mathbf{k}_{2}}+\sum_{\mathbf{k}_{3}}O_{\mathbf{k},\mathbf{k}_{3}}^{\text{}}\delta N_{\mathbf{k}_{3}}.
\end{equation}
Noting that $\left(\bar{N}_{\mathbf{k}}+1\right)=e^{\beta\varepsilon_{\mathbf{k}}}\bar{N}_{\mathbf{k}}$ and using momentum conservation, the constant shift $C$ is
\begin{align}
C\left[\left\{ \bar{N_{\mathbf{k}}},\bar{N_{\mathbf{k}_{1}}},\bar{N_{\mathbf{k}_{2}}},\bar{N_{\mathbf{k}_{3}}}\right\} \right] & =\frac{2\pi}{\hbar}\sum_{\mathbf{k}_{1},\mathbf{k}_{2},\mathbf{k}_{3}}
\delta\left(\varepsilon_{\mathbf{k}}+\varepsilon_{\mathbf{k}_{1}}-\varepsilon_{\mathbf{k}_{2}}-\varepsilon_{\mathbf{k}_{3}}\right)
\delta_{\mathbf{k}+\mathbf{k}_{1}-\mathbf{k}_{2}-\mathbf{k}_{3}}
\abs{W_{\mathbf{k},\mathbf{k}_{1},\mathbf{k}_{2},\mathbf{k}_{3}}^{+2}}^{2}\nonumber\\
 & \times\left[(\bar{N}_{\mathbf{k}} +1) (\bar{N}_{\mathbf{k}_{1}}+1)\bar{N}_{\mathbf{k}_{2}}\bar{N}_{\mathbf{k}_{3}}
 -\bar{N}_{\mathbf{k}} \bar{N}_{\mathbf{k}_{1}}\left(\bar{N}_{\mathbf{k}_{2}}+1\right)\left(\bar{N}_{\mathbf{k}_{3}}+1\right)\right]\nonumber\\
 & =0.
\end{align}
Here, $C$ is expected to be zero since there
should not be any finite current at equilibrium. The diagonal scattering
rate is 
\begin{align}
D_{\mathbf{k}} & =\frac{2\pi}{\hbar}\sum_{\mathbf{k}_{1},\mathbf{k}_{2},\mathbf{k}_{3}}
\delta\left(\varepsilon_{\mathbf{k}}+\varepsilon_{\mathbf{k}_{1}}-\varepsilon_{\mathbf{k}_{2}}-\varepsilon_{\mathbf{k}_{3}}\right)
\delta_{\mathbf{k}+\mathbf{k}_{1}-\mathbf{k}_{2}-\mathbf{k}_{3}}
\abs{W_{\mathbf{k},\mathbf{k}_{1},\mathbf{k}_{2},\mathbf{k}_{3}}^{+2}}^{2}
\frac{(\bar{N}_{\mathbf{k}_{1}}+1)\bar{N}_{\mathbf{k}_{2}}\bar{N}_{\mathbf{k}_{3}}}{\bar{N}_{\mathbf{k}}}
\end{align}
and the off-diagonal elements are 
\begin{align}
O_{\mathbf{k},\mathbf{k}_{1}}&=\frac{2\pi}{\hbar}\sum_{\mathbf{k}_{2},\mathbf{k}_{3}}\delta\left(\varepsilon_{\mathbf{k}}+\varepsilon_{\mathbf{k}_{1}}-\varepsilon_{\mathbf{k}_{2}}-\varepsilon_{\mathbf{k}_{3}}\right)
\delta_{\mathbf{k}+\mathbf{k}_{1}-\mathbf{k}_{2}-\mathbf{k}_{3}}
\abs{W_{\mathbf{k},\mathbf{k}_{1},\mathbf{k}_{2},\mathbf{k}_{3}}^{+2}}^{2}
\frac{(\bar{N}_{\mathbf{k}}+1)\bar{N}_{\mathbf{k}_{2}}\bar{N}_{\mathbf{k}_{3}}}{\bar{N}_{\mathbf{k}_{1}}}\\
O_{\mathbf{k},\mathbf{k}_{2}}&=\frac{2\pi}{\hbar}\sum_{\mathbf{k}_{1},\mathbf{k}_{3}}\delta\left(\varepsilon_{\mathbf{k}}+\varepsilon_{\mathbf{k}_{1}}-\varepsilon_{\mathbf{k}_{2}}-\varepsilon_{\mathbf{k}_{3}}\right)
\delta_{\mathbf{k}+\mathbf{k}_{1}-\mathbf{k}_{2}-\mathbf{k}_{3}}
\abs{W_{\mathbf{k},\mathbf{k}_{1},\mathbf{k}_{2},\mathbf{k}_{3}}^{+2}}^{2}
\frac{(\bar{N}_{\mathbf{k}}+1)(\bar{N}_{\mathbf{k}_{1}}+1)\bar{N}_{\mathbf{k}_{3}}}{\bar{N}_{\mathbf{k}_{2}}}\\
O_{\mathbf{k},\mathbf{k}_{3}}&=\frac{2\pi}{\hbar}\sum_{\mathbf{k}_{1},\mathbf{k}_{3}}
\delta\left(\varepsilon_{\mathbf{k}}+\varepsilon_{\mathbf{k}_{1}}-\varepsilon_{\mathbf{k}_{2}}-\varepsilon_{\mathbf{k}_{3}}\right)
\delta_{\mathbf{k}+\mathbf{k}_{1}-\mathbf{k}_{2}-\mathbf{k}_{3}}
\abs{W_{\mathbf{k},\mathbf{k}_{1},\mathbf{k}_{2},\mathbf{k}_{3}}^{+2}}^{2}
\frac{(\bar{N}_{\mathbf{k}}+1)(\bar{N}_{\mathbf{k}_{1}}+1)\bar{N}_{\mathbf{k}_{2}}}{\bar{N}_{\mathbf{k}_{3}}}.
\end{align}
To examine if the off-diagonal components contribute to $\kappa_{\mathrm{th,H}}$,
we rewrite $O_{\mathbf{k,k'}}$ in Hardy's basis:
\begin{align}
\mathcal{O}_{\mathbf{k},\mathbf{k}'}=\frac{\sqrt{\left(\bar{N}_{\mathbf{k}'}+1\right)\bar{N}_{\mathbf{k}'}}}{\sqrt{\left(\bar{N}_{\mathbf{k}}+1\right)\bar{N}_{\mathbf{k}}}}O_{\mathbf{k},\mathbf{k}'}.
\end{align}
As an example, we examine $\mathcal{O}_{\mathbf{k,k}_{1}}$ in Hardy's basis, given by 
\begin{align}
\mathcal{O}_{\mathbf{k},\mathbf{k}_{1}}=\frac{2\pi}{\hbar}\sum_{\mathbf{k}_{2},\mathbf{k}_{3}}\delta\left(\varepsilon_{\mathbf{k}}+\varepsilon_{\mathbf{k}_{1}}-\varepsilon_{\mathbf{k}_{2}}-\varepsilon_{\mathbf{k}_{3}}\right)
\delta_{\mathbf{k}+\mathbf{k}_{1}-\mathbf{k}_{2}-\mathbf{k}_{3}}
\abs{W_{\mathbf{k},\mathbf{k}_{1},\mathbf{k}_{2},\mathbf{k}_{3}}^{+2}}^{2}
\frac{\sqrt{\left(\bar{N}_{\mathbf{k}_{1}}+1\right)\left(\bar{N}_{\mathbf{k}}+1\right)}}{\sqrt{\bar{N}_{\mathbf{k}}\bar{N}_{\mathbf{k}_{1}}}}\bar{N}_{\mathbf{k}_{2}}\bar{N}_{\mathbf{k}_{3}}.
\end{align}
Since $\mathbf{k}$ and $\mathbf{k}_{1}$ play the same role, swapping these will not change the vertex. Thus, it is clear that $\mathcal{O}_{\mathbf{k},\mathbf{k}_{1}}-\mathcal{O}_{\mathbf{k_{1}},\mathbf{k}}=0$,
implying it doesn't contribute to $\kappa_{\mathrm{th,H}}$. A similar argument can be made for $\mathcal{O}_{\mathbf{k},\mathbf{k}_{2}}$ and $\mathcal{O}_{\mathbf{k},\mathbf{k}_{3}}$,
thus the first order diagrams in $\Delta N=0$ cannot generate a finite thermal Hall effect. Indeed, it was shown that all first order off-diagonal scattering processes for magnons~\cite{Chatzichrysafis2024} as well as phonons~\cite{Mangeolle2022PRB} do not contribute to transverse thermal transport. 

\subsubsection{Interference term}
As an example, we examine the interference between the tree level diagram and the bubble composed of 1-out-3-in (-3) and 1-in 3-out (+3) processes with $\Delta N = 0$. We have 

\begin{equation}
T_{\text{if}}^{(1),\text{out}}=W_{\mathbf{k}\mathbf{k}_{1}\mathbf{k}_{2}\mathbf{k}_{3}}^{+2}
\sqrt{N_{\mathbf{k}}N_{\mathbf{k}_{1}}(N_{\mathbf{k}_{2}}+1)(N_{\mathbf{k}_{3}}+1)}
\delta_{\mathbf{k}+\mathbf{k}_{1}-\mathbf{k}_{2}-\mathbf{k}_{3}}
\end{equation}
and
\begin{align}
T_{\text{if}}^{(2),\text{out}} & =\sqrt{N_{\mathbf{k}}N_{\mathbf{k}_{1}}(N_{\mathbf{k}_{2}}+1)(N_{\mathbf{k}_{3}}+1)}\times\nonumber\\
 & \sum_{\mathbf{p}_{1}\mathbf{p}_{2}}\frac{W_{\mathbf{k}_{1}\mathbf{p}_{1}\mathbf{p}_{2}\mathbf{k}}^{-3}W_{\mathbf{k}_{2}\mathbf{p}_{2}\mathbf{p}_{1}\mathbf{k}_3}^{+3}\left(N_{\mathbf{p}_{1}}+1\right)N_{\mathbf{p}_{2}}}{\varepsilon_{\mathbf{k}_{1}}+\varepsilon_{\mathbf{k}}-\varepsilon_{\mathbf{p}_{1}}+\varepsilon_{\mathbf{p}_{2}}+i\eta}
 \delta_{\mathbf{k}+\mathbf{k}_{1}+\mathbf{p}_{2},\mathbf{p}_{1}}
 \delta_{\mathbf{p}_{2}+\mathbf{k}_{2}+\mathbf{k}_3,\mathbf{p}_{1}}.
\end{align}
Let $\Delta E=\varepsilon_{\mathbf{k}_{1}}+\varepsilon_{\mathbf{k}}-\varepsilon_{\mathbf{p}_{1}}+\varepsilon_{\mathbf{p}_{2}}$.
The real and imaginary parts of $T_{\text{if}}^{(2),\text{out}}$
can be obtained using $\frac{1}{\Delta E+i\eta}=\mathcal{P}\left(\frac{1}{\Delta E}\right)-i\pi\delta\left(\Delta E\right).$
Putting everything together, the total scattering rate for the 2-in-2-out
tree $\times$ 3-in 1-out / 1-in 3-out bubble is 
\begin{align}
\Gamma_{\mathbf{k}}^{\left(1,2\right),\text{out}} & =\frac{4\pi}{\hbar}\frac{1}{2}\sum_{\mathbf{k}_{1}\mathbf{k}_{2}\mathbf{k}_{3}}
N_{\mathbf{k}}N_{\mathbf{k}_{1}}(N_{\mathbf{k}_{2}}+1)(N_{\mathbf{k}_{3}}+1)
\delta_{\mathbf{k}+\mathbf{k}_{1}-\mathbf{k}_{2}-\mathbf{k}_{3}}
\delta\left(\varepsilon_{\mathbf{k}}+\varepsilon_{\mathbf{k}_{1}}-\varepsilon_{\mathbf{k}_{2}}-\varepsilon_{\mathbf{k}_{3}}\right)\times\nonumber\\
 & \sum_{\mathbf{p}_{1}\mathbf{p}_{2}}\left(N_{\mathbf{p}_{1}}+1\right)N_{\mathbf{p}_{2}}\delta_{\mathbf{k}+\mathbf{k}_{1}+\mathbf{p}_{2},\mathbf{p}_{1}}
 \delta_{\mathbf{p}_{2}+\mathbf{k}_{2}+\mathbf{k}_3,\mathbf{p}_{1}}
 \left[\mathcal{R}_{\text{S}}^{\text{out}}+\mathcal{R}_{\text{AS}}^{\text{out}}\right],
\end{align}
where the extra factor of $1/2$ comes from separating the expression
into its symmetric and antisymmetric components 
\begin{align}
\mathcal{R}_{\text{S}}^{\text{out}} & =\mathcal{P}\left(\frac{1}{\Delta E}\right)\left[\text{Re}\left(W_{+2}\right)A+\text{Im}\left(W_{+2}\right)B^{\text{}}\right]+\pi\delta\left(\Delta E\right)\left[\text{Re}\left(W_{+2}\right)B-\text{Im}\left(W_{+2}\right)A\right]\text{ and}\nonumber\\
\mathcal{R}_{\text{AS}}^{\text{out}} & =\mathcal{P}\left(\frac{1}{\Delta E}\right)\left[\text{Re}\left(W_{+2}\right)A-\text{Im}\left(W_{+2}\right)B\right]+\pi\delta\left(\Delta E\right)\left[\text{Re}\left(W_{+2}\right)B+\text{Im}\left(W_{+2}\right)A\right].
\end{align}
For brevity, in the above we have denoted $W_{\mathbf{k}\mathbf{k}_{1}\mathbf{k}_{2}\mathbf{k}_{3}}^{+2}\equiv W_{+2}$,
$W_{\mathbf{k}_{1}\mathbf{p}_{1}\mathbf{p}_{2}\mathbf{k}}^{-3}\equiv W_{-3}$,
and $W_{\mathbf{k}_{2}\mathbf{p}_{2}\mathbf{p}_{1}\mathbf{k}_3}^{+3}\equiv W_{+3}$.

Similarly, the scattering rate for the in-process is given by 
\begin{align}
\Gamma_{\mathbf{k}}^{\left(1,2\right),\text{in}} & =\frac{4\pi}{\hbar}\frac{1}{2}\sum_{\mathbf{k}_{1}\mathbf{k}_{2}\mathbf{k}_{3}}
(N_{\mathbf{k}} +1) (N_{\mathbf{k}_{1}}+1)N_{\mathbf{k}_{2}}N_{\mathbf{k}_{3}}
\delta\left(\varepsilon_{\mathbf{k}}+\varepsilon_{\mathbf{k}_{1}}-\varepsilon_{\mathbf{k}_{2}}-\varepsilon_{\mathbf{k}_{3}}\right)
\delta_{\mathbf{k}+\mathbf{k}_{1}-\mathbf{k}_{2}-\mathbf{k}_{3}}\times\nonumber\\
 & \sum_{\mathbf{p}_{1}\mathbf{p}_{2}}N_{\mathbf{p}_{2}}\left(N_{\mathbf{p}_{1}}+1\right)
 \delta_{\mathbf{k}+\mathbf{k}_{1}+\mathbf{p}_{2},\mathbf{p}_{1}}\delta_{\mathbf{p}_{2}+\mathbf{k}_{2}+\mathbf{k}_3,\mathbf{p}_{1}}\left[\mathcal{R}_{\text{S}}^{\text{in}}+\mathcal{R}_{\text{AS}}^{\text{in}}\right],
\end{align}
where $\mathcal{R}_{\text{S}}^{\text{in}}=\left(\mathcal{R}_{\text{S}}^{\text{out}}\right)^{*}$
and $\mathcal{R}_{\text{AS}}^{\text{in}}=\left(\mathcal{R}_{\text{AS}}^{\text{out}}\right)^{*}$.
Next, we perform the same procedure of linearizing and transforming
into Hardy's basis. In this case, the off-diagonal components always correspond to $\mathcal{O}_{\mathbf{kk'}}^{++}$ or $\mathcal{O}_{\mathbf{kk'}}^{--}$ type processes, so the terms contributing to $\kappa_{\mathrm{th,H}}$ will be
those who break $\mathcal{O_{\mathbf{kk'}}^{++}}/\mathcal{O_{\mathbf{kk'}}^{--}}=e^{-\beta\varepsilon_{\mathbf{k'}}}$. Using the fact that $\text{Im}\left(W\right)=-\text{Im}\left(W^{*}\right)$,
and that $\left(\bar{N}_{\mathbf{k}}+1\right)=e^{\beta\varepsilon_{\mathbf{k}}}\bar{N}_{\mathbf{k}}$,
we have 

\begin{align}
\mathcal{O}_{\mathbf{kk}_{1}}^{++} & =\frac{4\pi^{2}}{\hbar}\frac{1}{2}\sum_{\mathbf{k}_{2}\mathbf{k}_{3}}\delta\left(\varepsilon_{\mathbf{k}}+\varepsilon_{\mathbf{k}_{1}}-\varepsilon_{\mathbf{k}_{2}}-\varepsilon_{\mathbf{k}_{3}}\right)
\delta_{\mathbf{k}+\mathbf{k}_{1}-\mathbf{k}_{2}-\mathbf{k}_{3}}
\bar{N}_{\mathbf{k}_{1}}\bar{N}_{\mathbf{k}_{2}}\bar{N}_{\mathbf{k}_{3}}\sqrt{e^{\beta(\varepsilon_{\vec{k}_1}+\varepsilon_{\vec{k}})}}\nonumber\\
 & \qquad\times\sum_{\mathbf{p}_{1}\mathbf{p}_{2}}\left(\bar{N}_{\mathbf{p}_{1}}+1\right)\bar{N}_{\mathbf{p}_{2}}\delta_{\mathbf{k}+\mathbf{k}_{1}+\mathbf{p}_{2},\mathbf{p}_{1}}\delta_{\mathbf{p}_{2}+\mathbf{k}_{2}+\mathbf{k}_3,\mathbf{p}_{1}}
 \delta\left(\varepsilon_{\mathbf{k}_{1}}+\varepsilon_{\mathbf{k}}-\varepsilon_{\mathbf{p}_{1}}+\varepsilon_{\mathbf{p}_{2}}\right)\nonumber\\
 & \qquad\times(-\text{Im}\left(W_{+2}\right)\left[\text{Re}\left(W_{+3}\right)\text{Re}\left(W_{-3}\right)-\text{Im}\left(W_{+3}\right)\text{Im}\left(W_{-3}\right)\right]\nonumber\\
 & \qquad -\text{Re}\left(W_{+2}\right)\left[\text{Re}\left(W_{+3}\right)\text{Im}\left(W_{-3}\right)-\text{Im}\left(W_{+3}\right)\text{Re}\left(W_{-3}\right)\right])\label{eq:okk1}\\
 & =-e^{-\beta\varepsilon_{\mathbf{k}_{1}}}\mathcal{O}_{\mathbf{kk}_{1}}^{--}
\end{align}
which obeys anti-detailed balance. A similar situation arises for
$\mathcal{O}_{\mathbf{kk}_{2}}$ and $\mathcal{O}_{\mathbf{kk}_{3}}$.
Thus, the thermal Hall conductivity is given by 
\begin{align*}
\kappa_{\mathrm{th,H}} & =\frac{1}{4k_{B}T^{2}V}\sum_{\mathbf{k,k'}}\mathbf{v}_{\mathbf{k}}\times\mathbf{v}_{\mathbf{k}'}\varepsilon_{\mathbf{k}}\varepsilon_{\mathbf{k}'}\tau_{\mathbf{k}}\tau_{\mathbf{k'}}G_{\mathbf{k}}G_{\mathbf{k'}}\frac{\bar{N}_{\mathbf{k}'}-\bar{N}_{\mathbf{k}}}{\bar{N}_{\mathbf{k}'}}\left(\mathcal{O}_{\mathbf{kk'}}^{++}-e^{-\beta\varepsilon_{\mathbf{k}'}}\mathcal{O}_{\mathbf{kk'}}^{--}\right)\\
& =\frac{1}{2k_{B}T^{2}V}\sum_{\mathbf{k,k'}}\mathbf{v}_{\mathbf{k}}\times\mathbf{v}_{\mathbf{k}'}\varepsilon_{\mathbf{k}}\varepsilon_{\mathbf{k}'}\tau_{\mathbf{k}}\tau_{\mathbf{k'}}G_{\mathbf{k}}G_{\mathbf{k'}}\frac{\bar{N}_{\mathbf{k}'}-\bar{N}_{\mathbf{k}}}{\bar{N}_{\mathbf{k}'}}\mathcal{O}_{\mathbf{kk'}}^{++}.
\end{align*}
From Eq.~\ref{eq:okk1}, it is clear that only odd powers of $J_\chi$ may contribute something finite to $\kappa_{\mathrm{th,H}}$, since the vertices from $J_1$ and $J_2$ are purely real, while those from $J_\chi$ are purely imaginary. To the lowest order of $J_\chi$, $\kappa_{\mathrm{th,H}}\propto J_1^2 J_\chi / S^2 \propto \sin(\pi \Phi/\Phi_0)$. In this case, the behaviour of $\kappa_{\mathrm{th,H}}$ reflects the fact that the $J_\chi$ interaction in the Hamiltonian is TR odd.

\section{Supplemental Plots}

\begin{figure}[htbp]
    \centering
    \includegraphics[width=\linewidth]{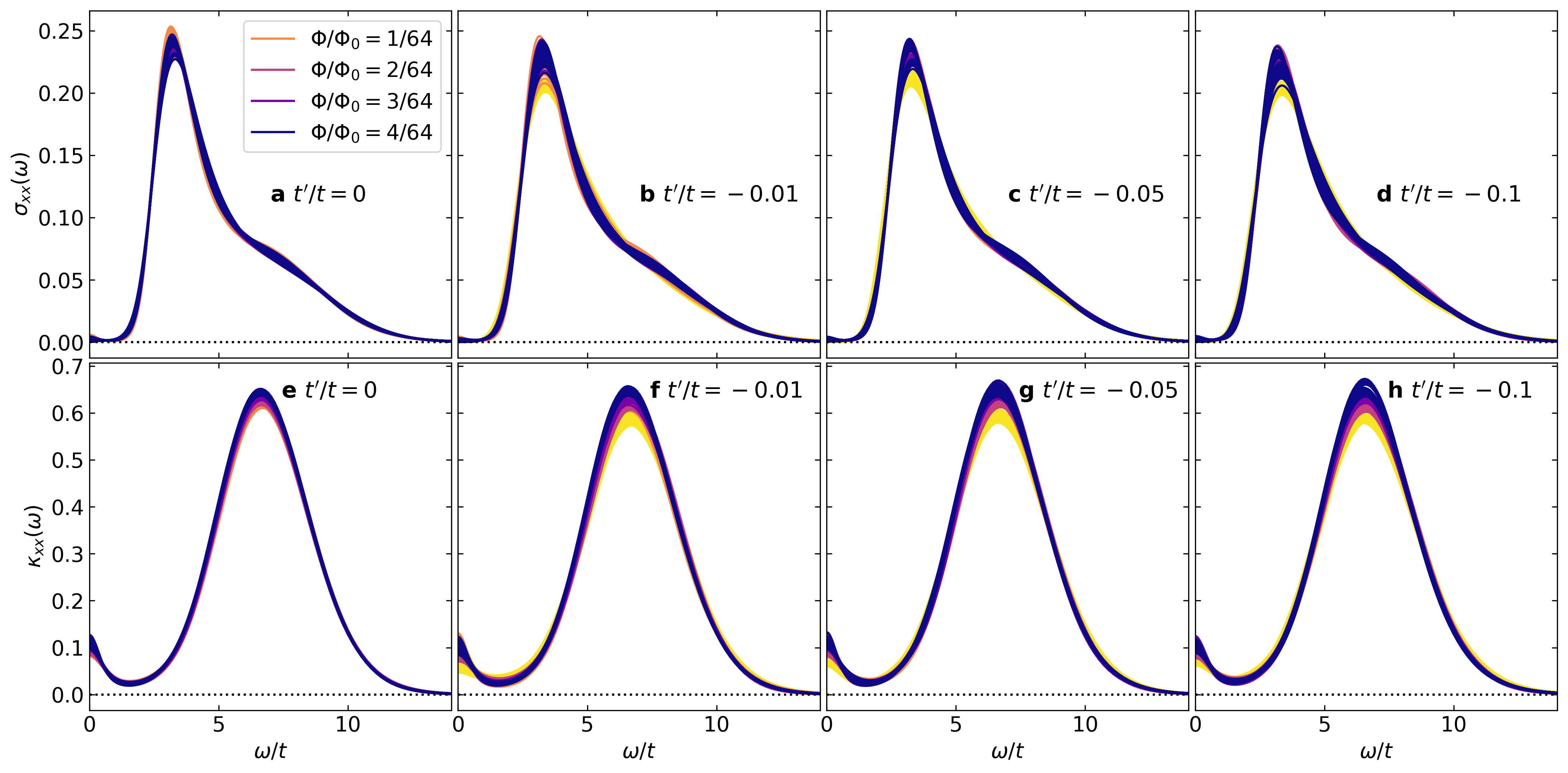}
    \caption[Minimal field and $t'/t$ dependence of longitudinal conductivities]{Example of minimal magnetic field strength and $t'/t$ dependence of longitudinal optical conductivity $\sigma_{xx}(\omega)$ and longitudinal frequency-dependent thermal conductivity $\kappa_{xx}(\omega)$. Temperature $\beta t = 5$, Hubbard $U/t = 6$. All panels share the same legend.}
    \label{fig:optical-long-var-tp-var-B}
\end{figure}

\begin{figure}[htbp]
    \centering
    \includegraphics[width=0.8\linewidth]{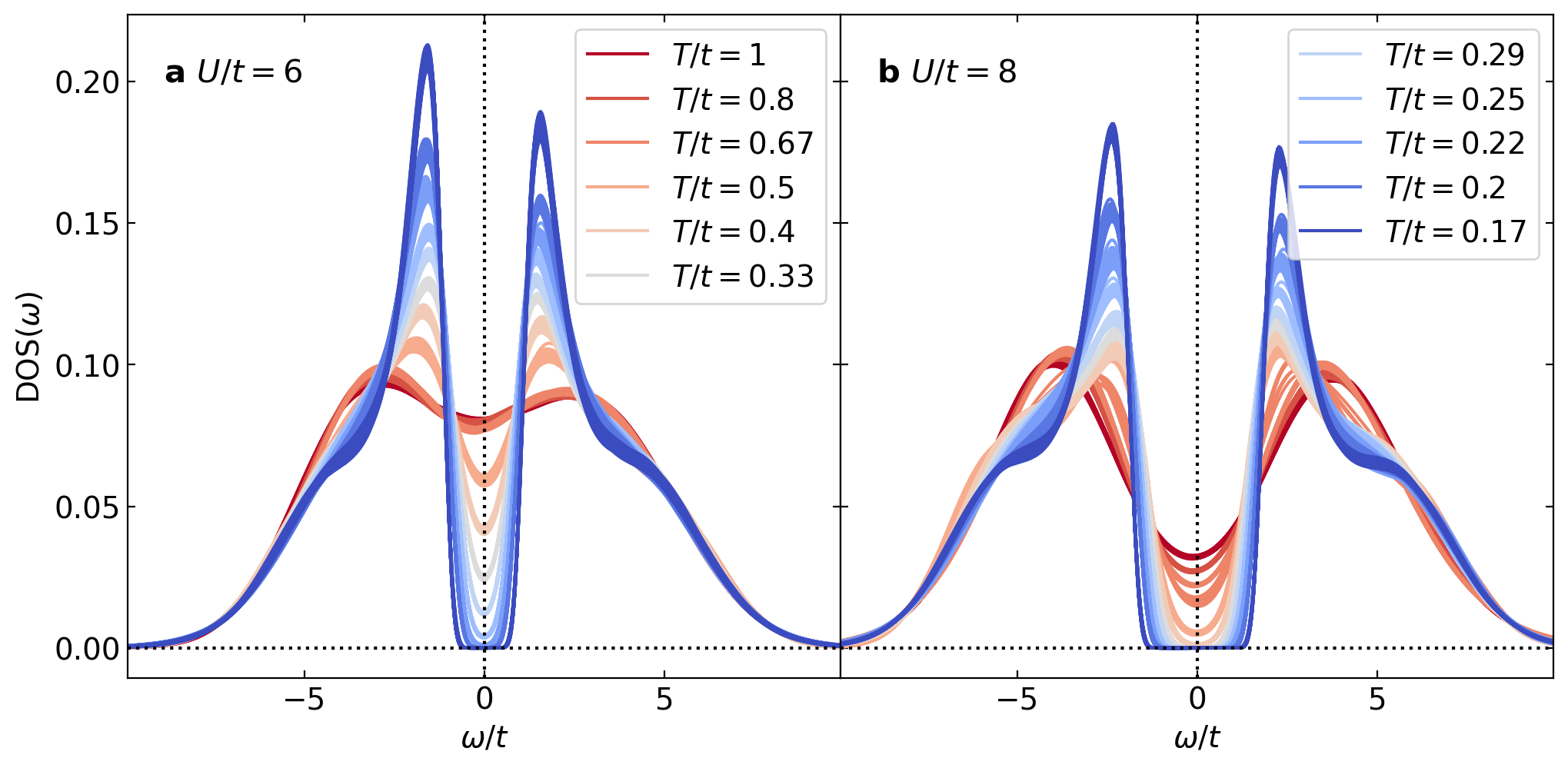}
    \caption[Temperature dependence of density of states]{Temperature and Hubbard $U$ dependence of density of states at \textbf{a} $U/t = 6$ and \textbf{b} $U/t=8$ for the Hubbard-Hofstadter model with $t'/t = -0.1$ and fixed field strength $\Phi/\Phi_0 = 1/64$ at half filling $\langle n \rangle = 1$. 100 bootstrap resamples are shown. Both panels share the same legend.}
    \label{fig:DOS-UT-dep}
\end{figure}

\begin{figure}[htbp]
    \centering
    \includegraphics[width=0.7\linewidth]{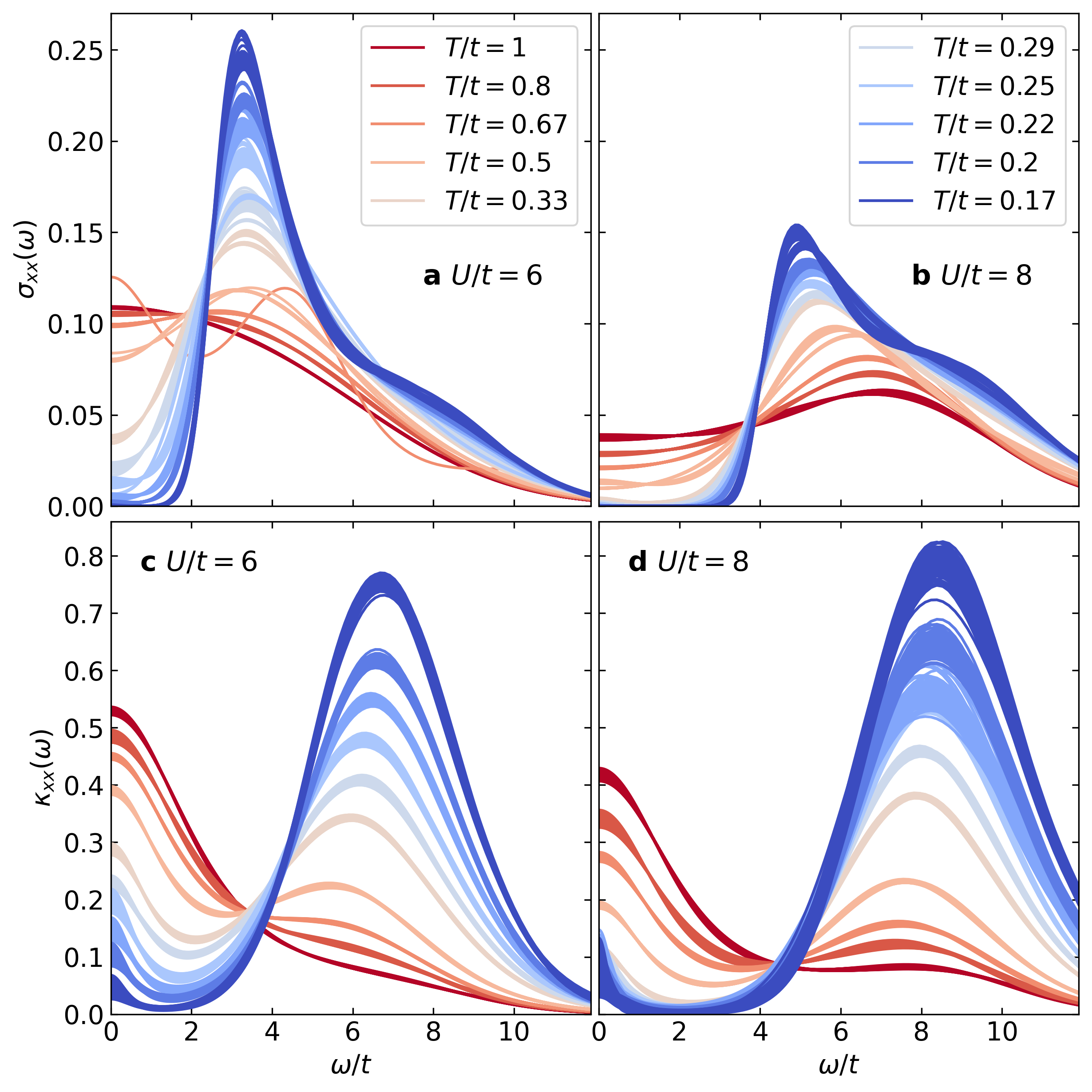}
    \caption[Temperature and U dependence longitudinal conductivities]{Temperature and Hubbard $U$ dependence of longitudinal frequency-dependent \textbf{a} \textbf{c} electrical conductivity $\sigma_{xx}(\omega)$, and \textbf{b} \textbf{d} thermal conductivity $\kappa_{xx}(\omega)$, for the Hubbard-Hofstadter model with $t'/t = -0.1$ at half-filling $\langle n \rangle = 1$ and fixed field strength $\Phi/\Phi_0 = 1/64$. 100 bootstrap resamples are shown. All panels share the same legend.}
    \label{fig:s-k-UT-dep}
\end{figure}

\begin{figure}[htbp]
    \centering
    \includegraphics[width=0.7\columnwidth]{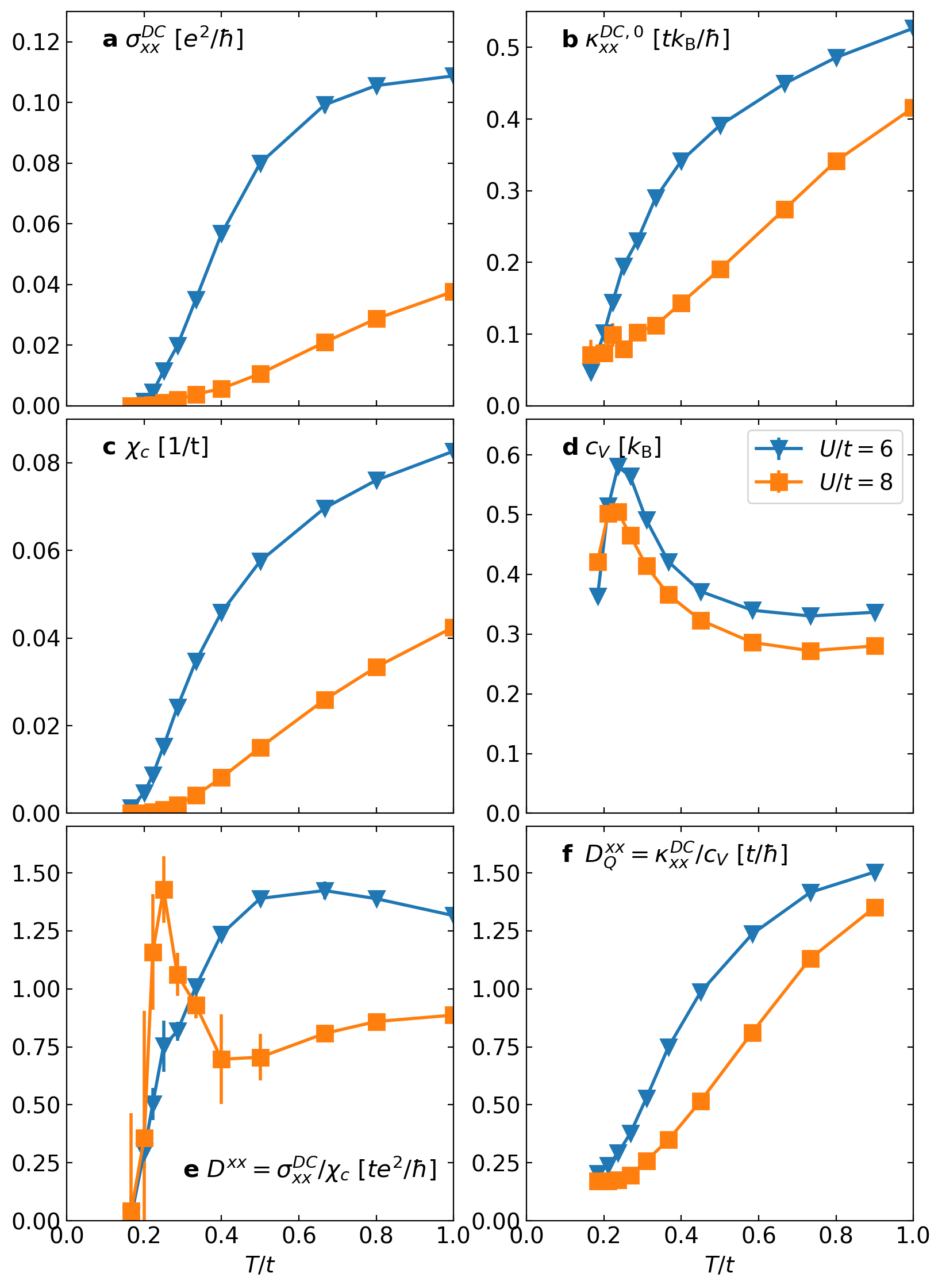}
    \caption{\textbf{a} Longitudinal DC electrical conductivity $\sigma_{xx}^{DC}$, \textbf{b} longitudinal DC thermal conductivity $\kappa_{xx}^{DC}$, \textbf{c} charge compressibility $\chi_c$, \textbf{d} specific heat $c_V$, \textbf{e} charge diffusivity $D^{xx}$, and \textbf{f} thermal diffusivity $D_Q^{xx}$ in the Hubbard-Hofstadter model with $t'/t=-0.1$ at half-filling $\langle n \rangle = 1$ and magnetic field strength $\Phi/\Phi_0 = 1/64$. All panels share the same legend.}
    \label{fig:DC-T-dep-var-U}
\end{figure}

\begin{figure}[htbp]
    \centering
    \includegraphics[width=\columnwidth]{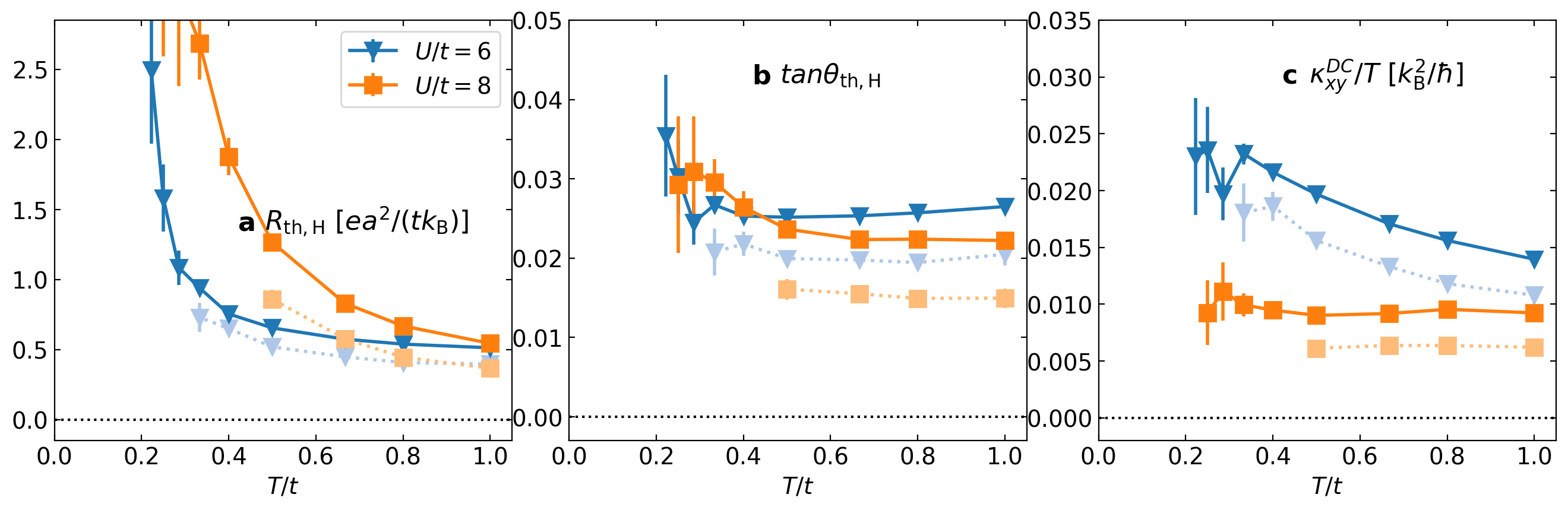}
    \caption[DC thermal Hall coefficient, thermal Hall angle, and thermal Hall conductivity]{\textbf{a} DC thermal Hall coefficient, \textbf{b} thermal Hall angle, and \textbf{c} thermal Hall conductivity for Hubbard $U/t=6$ and $U/t=8$. Solid lines denote results obtained by the proxy method, while dotted lines denote results obtained by the subtraction method. Next-nearest neighbor hopping $t'/t = -0.1$ and field strength $\Phi/\Phi_0 = 1/64$. All panels share the same legend. 
    }
    \label{fig:proxy-U-dep}
\end{figure}

\begin{figure}[htbp]
    \centering
    \includegraphics[width=\columnwidth]{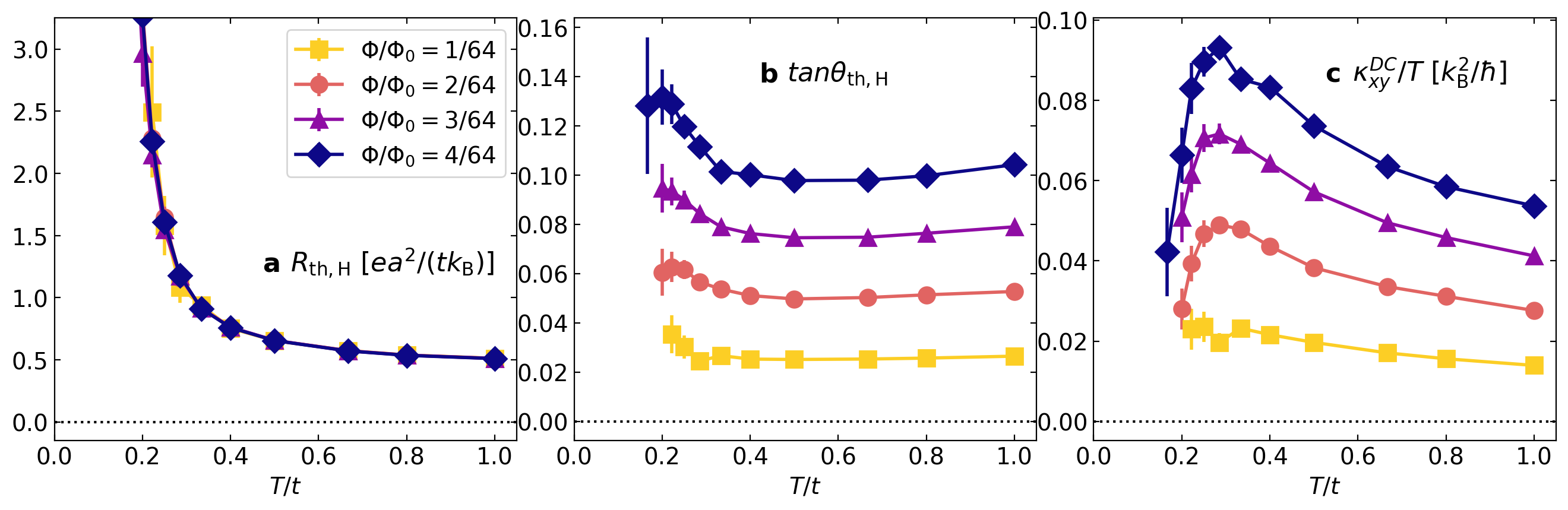}
    \caption[Magnetic field dependence of thermal Hall coefficient, thermal Hall angle, and thermal Hall conductivity]{Magnetic field dependence of \textbf{a} DC thermal Hall coefficient, \textbf{b} thermal Hall angle, and \textbf{c} thermal Hall conductivity. Results are obtained by the proxy method. Next-nearest neighbor hopping $t'/t = -0.1$, Hubbard interaction strength $U/t$. All panels share the same legend. 
    }
    \label{fig:proxy-B-dep}
\end{figure}

\bibliographystyle{unsrt}
\bibliography{main}